\documentclass[
 reprint,
footinbib,
 amsmath,amssymb,
 aps,
]{revtex4-2}

\usepackage{graphicx}
\usepackage{dcolumn}
\usepackage{bm}
\usepackage{hyperref}
\usepackage{placeins}
\usepackage{amssymb}
\font\scripty=cmsy10
\def\sy{\fam2\scripty}

\DeclareMathOperator*{\argmax}{arg\,max}

\usepackage{color}

\def\b{{\tilde\alpha}}
\def\q{{\tilde{\bf x}}}

\def\s{{\tilde v }}
\def\R{R}

\def\phiC{\phi_{\mathcal{C}}}

\begin{document}

\preprint{}

\title{Environmental path-entropy and collective motion}
\author{Harvey L. Devereux}
\affiliation{Department of Mathematics, University of Warwick, Coventry CV4 7AL, UK}
\author{Matthew S. Turner}
\affiliation{Department of Physics and Centre for Complexity Science, University of Warwick, Coventry CV4 7AL, UK}
\affiliation{Department of Chemical Engineering, Kyoto University, Kyoto, 615-8510, Japan}

\date{\today}

\begin{abstract}
Inspired by the swarming or flocking of animal systems we study groups of agents moving in unbounded 2D space. Individual trajectories derive from a ``bottom-up'' principle: individuals reorient to maximise their future path entropy over environmental states. This can be seen as a proxy for {\it keeping options open}, a principle that may confer evolutionary fitness in an uncertain world.
We find an ordered (co-aligned) state naturally emerges, as well as disordered states or rotating clusters; similar phenotypes are observed in birds, insects and fish, respectively. The ordered state exhibits an order-disorder transition under two forms of noise: (i) standard additive orientational noise, applied to the post-decision orientations (ii) ``cognitive'' noise, overlaid onto each individual's model of the future paths of other agents.
Unusually, the order increases at low noise, before later decreasing through the order-disorder transition as the noise increases further.
\end{abstract}

\maketitle
Collective motion occurs in both living and synthetic systems. In living systems this arises in a wide variety of species over different length scales, e.g. micro-organisms, cells, insects, fish, birds \cite{Sokolov2007Concentration,vicsek2012collective,Cavagna201005766,flack2018local,Zitterbart2011CoordinatedMP,Moussa} and even dinosaurs \cite{dinosaurs}. Interest in the physics community often lies in developing models of collective motion that are analogous to living systems, many of which exhibit ordered (coaligned) motion and support a noise-induced transition to disorder
\cite{Vicsek1995Novel,Buhl2006Recent,Szabo2006Phase,Gregoire2004,Chate2008,
PhysRevLettMetricFree,Weber2013Long,Calovi2014SwarmingSM}.
Long-ranged behavioural interactions may arise in nature and there have been some attempts to analyse such interactions
 \cite{Nishiguchi2017Long,PhysRevLettMetricFree,Ballerini1232Topological,starlingdisplays,Pearce10422,Pawel2020}. 
 These can naturally be traced to the nature of information transfer between agents 
 \cite{Gallup7245humanvisinfo,attanasi2014Information}, noting that senses like vision are long ranged. 
 
 Other models of swarming behaviour incorporate explicit alignment, cohesion, and/or collision avoidance rules directly into an agent-based model \cite{PhysRevLettMetricFree,Ballerini1232Topological,Toner1995Long}. However, such models cannot easily explain the underlying reason for the emergence
of properties like cohesion and coalignment as these are essentially incorporated into the models at the outset.
One recent alternative approaches is to utilise machine learning based on using a simple form of perception to maintain cohesion directly \cite{durve}. Another involves the study of large deviations of non-aligning active particles that are biased, e.g. by effective alignment of self-propulsion with particle velocity \cite{j1,j2,j3,j4}.

 While it is possible neural circuitry of animals encodes an algorithm that {\it directly} 
  targets coalignment and cohesion in the same mathematical manner as in these models
  it, seems much more likely that some lower-level principle is involved. 
  This principle, almost certainly associated with evolutionary fitness in some way, 
  might then be the origin of cohesion and coalignment. We argue that more satisfactory explanations 
  for the phenomenon of swarming may be offered by testing candidates for this lower level principle. 
  In this letter we analyse one such model.

There is a small but growing literature focussing on the causal understanding of complex behaviour, 
cast as an entropy or state maximisation approach. 
Here some measure of variation across \textit{future} paths accessible from a particular system configuration 
is computed and an action that maximises this variation is selected, e.g. 
\cite{wissner2013causal,empowermentRL,Mann20150037,Charlesworth15362,hornischer2019intelligence}.
It is argued that agents that can retain access to the most varied future environments can better select from 
these to satisfy any immediate requirements or objectives, e.g. resource acquisition or predator evasion. 
For these reasons such strategies are expected to generally confer evolutionary fitness in an uncertain world. The present work shares similar motivation to \cite{Charlesworth15362} but provides a rigorous mathematical model based on path-entropies and focuses on the emergence of order. We believe that such models offer clear advantages in terms of their conceptual clarity and prospects for future development.
\begin{figure}[ht!]
  \centering
  \includegraphics[width=\columnwidth]{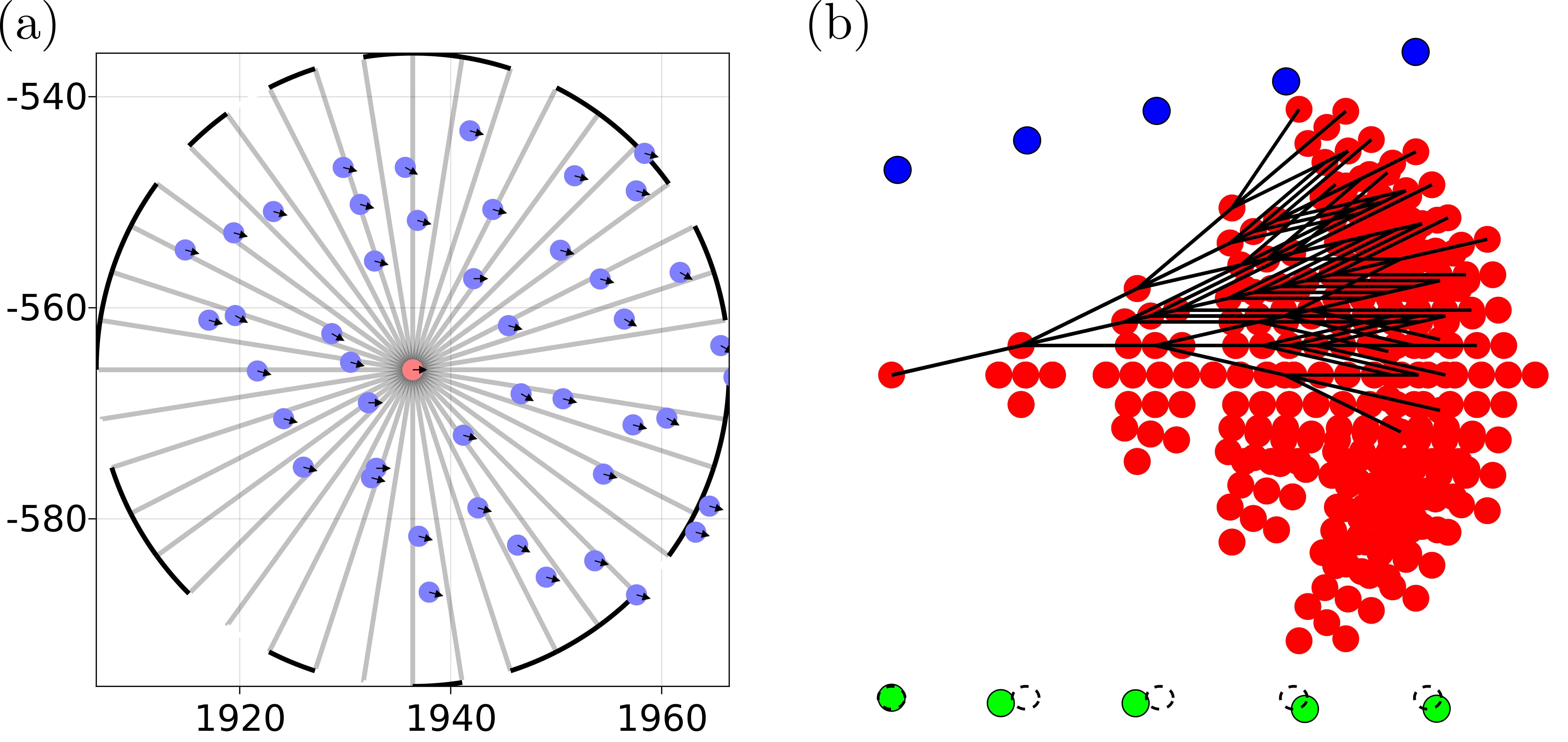}
  \caption{Snapshot of a system configuration and sketch explaining our model. 
   (a) $N=50$ agents that take actions
    to maximise a future path entropy over environmental states (see text); 
    axes show $x$-$y$ coordinates in units of the agent radius. 
    Overlaid (broken circle) is a representation of the visual state 
    perceived by the red individual. Obtained from 
    a simulation with parameters $\tau=6$, $\Delta \theta =\pi/12= 15^{\circ}$, $v_{0}=10$, 
    and $\Delta v = 2$ (see text for details). (b) 
    In red the tree of hypothetical future actions the agent examines, starting from the present at the root on the far-left.
    Shown in blue and green (with dashed circles)
    are ballistic and noisey motion assumptions of other agents.
     }
  \label{fig:setup}
\end{figure}

To realise such a model here, agents are treated as oriented unit disks that move in discrete time $t$, 
defining our length and time units, respectively. The position of the $i^{\rm th}$ agent in the next time 
step is
\begin{equation}
{\bf x}_{i}^{t+1}={\bf x}_{i}^{t}+{\bf v}_{i}^{t+1}.\label{translate}
\end{equation}
At each discrete time step $t$ agents choose from $z=5$ velocities: 
either to move along their current orientation with one of three speeds: nominal $v_{0}$, fast $v_{0}+\Delta v$ or slow $v_{0}-\Delta v$; or to reorientate by an angle $\pm \Delta \theta$ while moving at the nominal speed $v_0$. Unless noted otherwise we take $v_0=10$, $\Delta v=2$ and $\Delta \theta=\pi/12=15^\circ$.
The agent's velocity is updated by an operator $A_{\alpha_{i}^{t}}$ 
acting on its previous velocity ${\bf v}_i^{t}$
\begin{equation}
{\bf v}_i^{t+1}=A_{\alpha_{i}^{t}}[{\bf v}^{t}_{i}].\label{A1}
\end{equation}
Actions $\alpha$ change the velocity according to
\begin{equation}
A_\alpha[{\bf v}]=v_\alpha \R (\theta_\alpha){\bf \hat v}.\label{A2}
\end{equation}
The index $\alpha\in[1,z]$ labels possible actions, here with indices dropped for clarity. Hat accents denote unit vectors according to ${\bf \hat v}={\bf v}/|{\bf v}|$ throughout, with $|\cdot |$ the Euclidean norm. 
The action chosen at each time step determines the corresponding speed of the agent 
$v_1=v_4=v_5=v_{0}$, $v_2=v_{0}+\Delta v$, $v_3=v_{0}-\Delta v$ in that time step. Where $\R$
generates a rotation
\begin{align}
    \R(\theta) = \left(\begin{array}{cc}
        \cos\theta & \sin\theta \\
        -\sin\theta & \cos\theta
    \end{array}\right), \label{R}
\end{align}
with rotation angles $\theta_1=\theta_2=\theta_3=0$, $\theta_4=\Delta \theta$ and $\theta_5=-\Delta \theta$. 
The sequence of such actions realised by this agent $\alpha_{i}^{t}$ over time $t$ completely determine the dynamics.
In order to select actions, i.e compute hypothetical path entropy over future states, this model requires that agents model positions of themselves and other agents into the future. Therefore we adopt the notation $\q_{k}^{t'}$, $\b_{k}^{t'}$, $v_{\b_{k}^{t'}}$ and $\theta_{\b_{k}^{t'}}$,  
involving a tilde  accent, to indicate virtual positions, actions, speeds and rotation angles of  all agents $k$ at time $t'$.
Hence
\begin{align}
   \q_{k}^{t+s} &={\bf x}_{k}^{t}+\sum_{t'=t}^{t-1+s}\s_{\b_{k}^{t'}}\prod_{t''=t}^{t'}\R(\theta_{\b_{k}^{t''}})\hat{\bf v}_{k}^{t},\label{eq:futureState}
\end{align}
with $1\le s\le\tau$ reflecting the time horizon $\tau$.

Equation (\ref{eq:futureState}) generates the hypothetical position of both the $k=i$ (self) and $k=j\ne i$ (other) agents.
However, we make a simplifying assumption for the motion of the $j\ne i$ (other) agents.
Here our default model corresponds to ``ballistic'' translation of the $j\ne i$ agents in which 
$v_{\b_{j}^{t'}}=v_{0}$ and $\theta_{\b_{j}^{t'}}=0$, $\forall t'\ge t$. The speeds and rotations depend neither 
on the particle index $j$ nor the future time index and so they can be stated in more condensed form simply as 
$\s=v_{0}$ and $\tilde \theta=\theta_{\tilde\alpha}=0$. The ballistic assumption 
can often be rather good, in the sense that the trajectories that are realised
can have a very high degree of orientational order and so the assumption is broadly self-consistent 
\footnote{It is not strictly self-consistent in the sense that the model for others (ballistic motion)
is different to that of self (path entropy-maximisation). It is unclear whether such models can ever be made 
rigorously self-consistent or whether they then represent uncomputable functions \cite{turing_1937,Godel}}.
Later in this article we consider models that generate different virtual actions for the $j\ne i$ 
agents that incorporate noise. See Fig~\ref{fig:setup}(b) for a sketch of this dynamical scheme.

The environmental state of an agent is assumed to be perceived using only vision, 
see Fig~\ref{fig:setup}(a). This state encodes information on the relative 
positions of the other agents in a manner that is broadly consistent with animal vision, 
abstracted to $d=2$ dimensions: visual sensing involves a radial projection of all other 
agents onto a circular sensor array at each agent. Loosely speaking, the radial projection registers 
0 ``white'' along lines of sight not intersecting agents, and 1 ``black'' along those that do. We discretise this
into an $n_s$-dimensional visual state vector $\boldsymbol{\psi}_i$, for angular sub regions of size $2\pi/n_s$.
This then resembles a spin state, e.g. $(0,1,0,0,1\cdots)$. 

Mathematically we use two indicator functions, first the distance of shortest approach along a line-of-sight
$\hat{\bf n}_i=\R(\chi){\bf {\hat v}}_i$ originating from the $i^{\rm th}$ agent, 
\begin{equation}
  I_{ij}=\Theta[1-|\q_{ij}\times \hat{\bf n}_i(\chi)|] \label{Iij}.
\end{equation}
Where the Heaviside function $\Theta[x]=1$ for $x\ge 0$ and $0$ otherwise, $\q_{ij}$ 
is the separation vector $\tilde{\bf x}_j-\tilde{\bf x}_i$, with $|\cdot|$ the Euclidean norm. 
Equation (\ref{Iij}) indicates an agent is 
visible along this line of sight in {\it either} direction from the $i^{\rm th}$
agent, i.e. along $\chi$ or $\chi+\pi$. We restrict to $\chi$ using the second indicator
\begin{equation}
  I'_{ij}=\Theta[\q_{ij}\cdot\hat{\bf n}_i(\chi)].
\end{equation}
The $n^{\rm th}$ component of the visual state vector $\boldsymbol\psi_{i}$ is then
\begin{equation}
{\psi}_{i}^{n}=\Theta\Bigg[\int_{\sigma_{n}}\Theta\Big[\sum_{j}I_{ij}(\chi)I'_{ij}(\chi)\Big]d\chi-\frac{\pi}{n_s}\Bigg], \label{psi_i_n}
\end{equation}
\begin{figure}[ht!]
  \centering
  \includegraphics[width=\columnwidth]{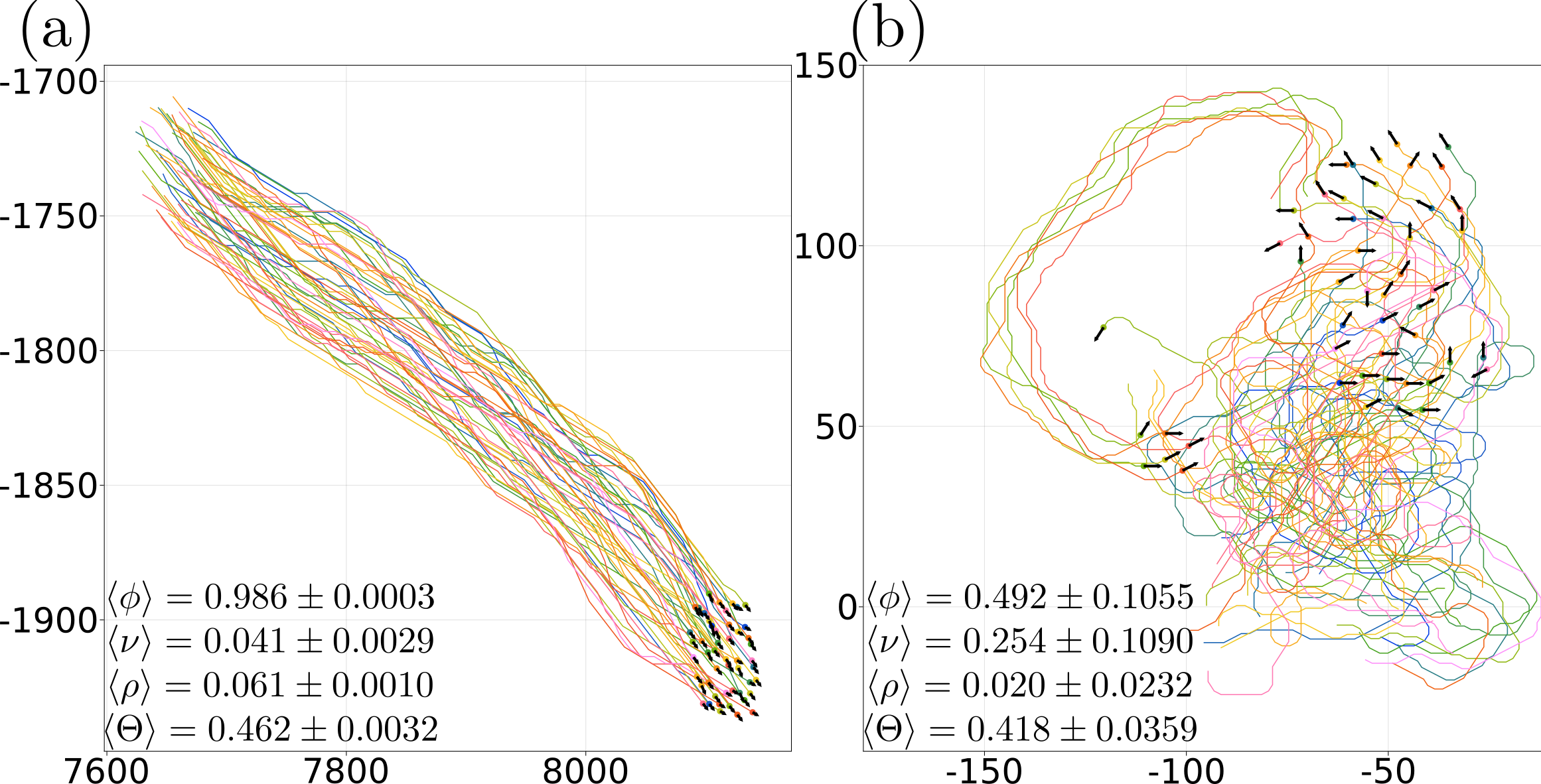}
  \caption{Agents that maximise environmental path-entropy naturally adopt different dynamical modes,
   or ``phenotypes''. 
   Each panel shows the agent's trajectories, 
   together with the time-averaged mean order $\phi$, 
   root-mean-squared vorticity $\nu$, density $\rho$ and opacity $\Theta$ (see text):
   (a) the ordered, dense (``bird'')  phenotype, 
   (b) translation combined with significant rotation (similar to ``fish'' or ``insects''). 
   Averages computed over 10 replicates.  }
  \label{fig:phenotype}
\end{figure}
where the $n^{\rm th}$ sensor covers the angular domain 
$\sigma_n=[2\pi(n-1)/n_{s},2\pi n/n_{s}]$. The inner Heaviside function registers 1 (``black'') if at least one
agent intersects line-of-sight $\chi$, the integral then measures the coverage of $\sigma_n$ by ``black'' regions. The outermost 
Heaviside function is a further threshold that at least half the sensor must be ``black'' to activate the $n^{\rm th}$ visual state
component. 

For a virtual action $\tilde\alpha_i^t$ the entropy of the state distribution over 
(all nodes on) all virtual paths for
the $i^{\rm th}$ agent following action $\tilde\alpha_i^t$ is
\begin{align}
    S(\tilde\alpha_i^t) = -\sum_{\boldsymbol{\psi}} p_i(\tilde\alpha_i^t,\boldsymbol{\psi})\log p_i(\tilde\alpha_i^t,\boldsymbol{\psi}). \label{eq:entropy}
\end{align}
Where $p_{i}(\tilde\alpha_i^t,\boldsymbol{\psi})$ is the count of occurences of a state $\boldsymbol{\psi}$ on these virtual paths,
normalised by the count of states on all branches. In this way each action-branch $\tilde\alpha_i^t$ is associated with an 
environmental path entropy $S(\tilde\alpha_i^t)$. The key step in the 
decision making process is that each agent then executes the action
\begin{align}
\alpha_i^t=\argmax_{\tilde\alpha_i^t} S{(\tilde\alpha_i^t)},
\label{eq:action}
\end{align}
thereby choosing the branch that maximises the entropy of future visual states. 
\begin{figure}[ht!]
  \centering
  \includegraphics[width=\columnwidth]{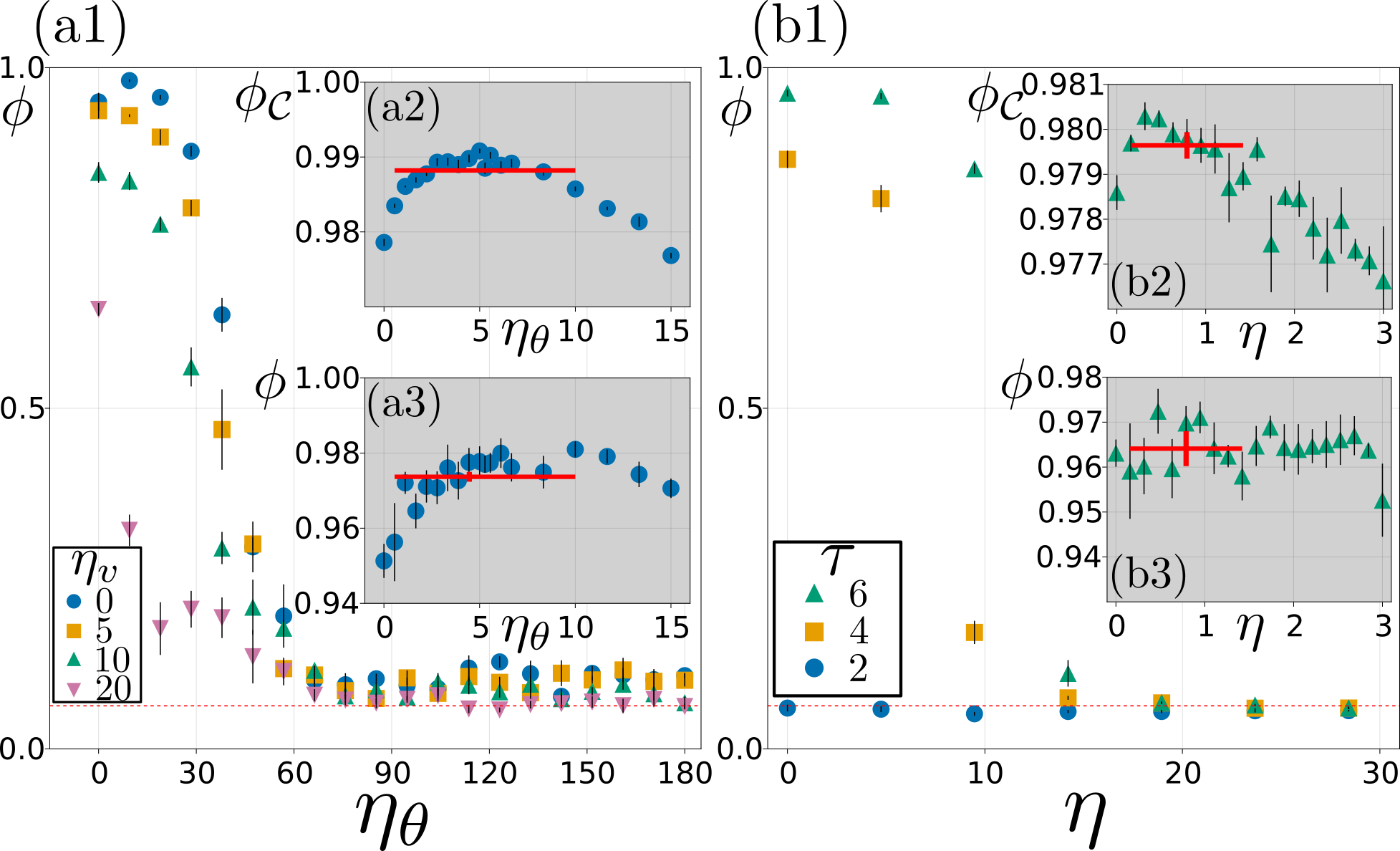}
  \caption{Ordering transitions in the presence of (a) cognitive noise and (b) post-decision 
  orientational noise: (a1-3) each agent approximates the future trajectories of others in the 
  presence of {\bf cognitive} noise as a sequence of random rotations and 
  speed changes from their current heading and speed $v_{0}$. The noise strengths $\eta_{\theta}$ 
  and $\eta_{v}$ characterise the magnitude of the rotations (degrees) and speed changes 
  (body radii per time step), respectively. (a1) A transition from high order to a disordered 
  phase occurs with increasing cognitive rotational noise $\eta_{\theta}$. Insets (a2) and (a3) 
  focus on small $\eta_{\theta}$ with $\eta_{v}=0$. They show the order of the largest cluster 
  $\phiC$ and the overall system order $\phi$, respectively; note the maximum in order appears 
  at non-zero noise. The red horizontal line shows the order averaged over all runs 
  $0<\eta\le 10^\circ$.  In (a1-3) the future time horizon is $\tau=6$. (b1) 
  shows the effect of {\bf post-decision} orientational noise on global order $\phi$. Here a 
  random rotation with root-mean
    squared angle $\eta$ (degrees) is applied directly to the velocity, before the movement 
    update. (b2) shows a statistically significant local maximum in
$\phiC$ for non-zero noise $\eta$ whereas (b3) now shows no significant maximum in $\phi$. 
The red horizontal line shows the order averaged over all runs $0<\eta\le 1.5^\circ$. 
All systems contain $N=250$; all error bars are 1 standard error in the mean; the dashed 
lines represent the mean order $\phi=1/\sqrt{N}$ of randomly orientated agents. In all statistical tests additional repeats (n=16) were computed for the zero noise case, see text for 
further details.}
\label{fig:noise}
\end{figure}
This process is carried out simultaneously for each agent and repeated, from scratch, at each time step. 
Degenerate options 
are selected at random, the only randomness in the baseline algorithm that is otherwise
deterministic.

Our model supports various phenotypes. In Fig~\ref{fig:phenotype} 
we report on the effect of varying the turning rate $\Delta \theta$, 
the nominal speed $v_{0}$ and its variation $\Delta v$. 
We find that a highly ordered and cohesive phenotype is commonly achieved 
when the agents move relatively fast with moderate turning. 
Resembling those seen in flocks of social birds \cite{BALLERINI2008201,HemelrijkPLOSONE2011}, 
noting that these birds also have relatively fast speed, do not slow significantly 
and have limited turning ability relative to an insect. 
We also find cohesive disordered groups, some showing circulation. 
The most important conditions for the emergence of cohesive swarms are (i) $\tau\gtrsim 3$, (ii) $10\lesssim n_s\lesssim 100$ to avoid the visual 
states becoming largely degenerate (see SI for details).

We report the visual opacity as the average sensor state $\Theta=\langle\frac1{n_s}\sum_{n=1}^{n_s} \psi_i^n\rangle$, density 
$\rho = \langle\frac{N\pi r^{2}}{{\sy A}^t}\rangle$ with the convex hull area ${\sy A}^t$, global order 
$\phi = \langle \bigl|\frac1N\sum_{i=1}^N\hat{\boldsymbol{v}}_{i}^t\bigr|\rangle$, and quantify rotation using a normalised mean squared 
vorticity $\nu^2 = \bigl\langle\bigl(\frac{1}{N}\sum_{i}\hat{\boldsymbol{r}}_{i}^t\times \hat{\boldsymbol{v}}_{i}^t\bigr)^{2}\bigr\rangle$, 
with $\boldsymbol{r}_{i}^t=\boldsymbol{x}_{i}^t-\langle \boldsymbol{x}_{k}^t\rangle_{k}$ and
$\boldsymbol{v}_{i}^t$ the $i^{\rm th}$ agent's position relative to the geometric centre and velocity respectively. In each case we average
over agents $i$ and times $t$. We also use 
a measure of spatial clustering using DBScan \cite{ester1996density} (SI section S2 for details) to both detect fragmentations and
measure the quantities above on clusters. We denote the average fraction of agents in the largest cluster as $\mathcal{C}$, and by $\phiC$ 
denote the order of the largest cluster.

We have established the emergence of co-aligned, cohesive states under environmental path entropy maximising trajectories, see Fig~\ref{fig:phenotype}(a). It is therefore natural to ask about the effect of noise on these dynamics. In this way we will investigate to what extent this model supports an order-disorder transition similar to those extensively studied in other models of collective motion \cite{Vicsek1995Novel,Buhl2006Recent,Szabo2006Phase,Gregoire2004,Chate2008,
PhysRevLettMetricFree,Weber2013Long,Calovi2014SwarmingSM}.

By ``cognitive noise'' we mean some imprecision in an agent's model of the others. We therefore define a stochastic process 
for the virtual speeds $\s^{t'}=v_{0}+\mu_v^{t'}$ and rotations $\tilde \theta^{t'}= \mu_\theta^{t'}$ of all $j\ne i$ agents. 
Here both $\mu$ variables (subscript $v$, $\theta$ omitted for clarity) are drawn from zero mean 
$\langle\mu^{t'}_j\rangle=0$ Gaussian distributions that are uncorrelated according to
$\langle \mu_j^{t'}\mu_{j'}^{t''}\rangle=\eta^2\delta_{jj'}\delta_{t't''}$ with $j,j'\ne i$
here the particle index of the (other) agents and $t',t''\ge t$. 
The root-mean-squared noise amplitude of the speed and orientation are written, with subscripts restored, 
as $\eta_v$ and  $\eta_\theta$ respectively. An example is shown in the sequence of 
positions shown in green in Fig~\ref{fig:setup}(b). All else proceeds as before, 
without any additional noise applied to realised agent actions.

The most striking feature from Fig~\ref{fig:noise}a is that the order initially {\it increases} 
with the addition of small levels of noise, before later decreasing again. 
An upper tailed t-test (with unequal variances) 
on the difference of the mean order in the noise-free case ($\eta_{\theta}=0$) and the mean of simulations with 
non-zero noise $0<\eta\leq 10^\circ$, rejects the null hypothesis that the mean order $\phi$ 
is the same at the level of $p<10^{-13}$. The same t-test for $\phiC$, 
the order computed only for agents that are members of the largest cluster, is rejected at $p<10^{-50}$.

To understand why a small amount of noise might actually increase order we compare the noise level at which the order is maximal,
roughly $\eta_\theta=5-7^\circ$, to intrinsic variation in the realised dynamics in a low noise state $\phi\approx0.98$. 
There are several ways to achieve this.
(i) approximating the order as the mean component of the normalised velocities of the agents along the average 
direction of motion, $\phi=\langle \cos \delta\theta\rangle\approx1-\frac12 \langle \delta\theta^2\rangle$, leading to $\delta\theta_{\rm rms}=11^\circ$ (ii)
Crudely assuming moves are uncorrelated extending for $\tau$ steps into the future and asking what angular noise amplitude {\it per time step}, analogous to $\eta_\theta$ would be required to give the realised 
order $\phi$ at the {\it end} of this sequence, leading to $11^\circ/\sqrt{\tau}=4.7^\circ$ (iii) using the velocity auto-correlation function 
$C_{vv}(t)=\langle \hat{{\bf v}}_{i}^{t'}\cdot \hat{\bf v}_{i}^{t'+t}\rangle$ (see SI Fig 4) and extracting an angular noise per time step by either 
writing $V_{vv}(1)=0.987=\langle \cos \delta\theta\rangle$ or by using $V_{vv}(\tau)\approx 0.968$ leading to $\delta\theta\approx 9^\circ$ 
and $6^\circ$ respectively. All are similar to the observed value of $\eta_\theta$ at which order is maximal. 
Thus the realised order is maximal at a value of cognitive noise $\eta_\theta$ that is {\it self-consistent} 
with the variation in the realised trajectories that arises in the dynamics. We argue that this is the noise level at 
which the predictive model of the trajectories of other agents will be more accurate, at least in a statistical sense. 
We propose that this represents the fundamental reason for the increase of order at small noise levels. 

To apply post-decision noise, the rotation associated with each action that appears in Eq~(\ref{A2}) 
is modified to include noise according to $\theta_1=\theta_2=\theta_3=\zeta_i^t$, $\theta_4=\Delta\theta+\zeta_i^t$ 
and $\theta_5=-\Delta\theta+\zeta_i^t$ with the random rotation angle $\zeta_i^t$ drawn from a zero mean 
$\langle\zeta_i^t\rangle=0$ Gaussian distribution satisfying
$\langle \zeta_i^t\zeta_j^{t'}\rangle=\eta^2\delta_{ij}\delta_{tt'}$. This noise can be interpreted as arising from imperfect 
implementation of the target velocity, ubiquitous in physical or living systems.

Figure \ref{fig:noise}b shows the effect of this post-decision orientational noise.
At large noise amplitude $\eta$ the order approaches a value $\sim{1}/{\sqrt{N}}$, expected 
for $N$ randomly orientated agents. This corresponds to a complete loss of orientational order.  
We find the order-disorder transition occurs around $\eta=12^{\circ}$. This is a significantly 
smaller noise level 
than for the case of cognitive noise, see Figure \ref{fig:noise} (a1-a3),
where the transition occurs around $\eta_\theta=45^{\circ}$. This indicates that 
cognitive noise has a much 
weaker disordering effect than post-decision noise and could even be seen as providing 
robustness, by anticipating 
the possibility of varied trajectories in the future. In contrast, post-decision noise 
plays no such role. 

A one-tailed t-test, to test whether the order at 
non-zero noise values are significantly different from the zero noise case, 
was performed for both $\phi$ and $\phiC$.
The result being significant for the mean order for agents 
in the largest cluster $\phiC$ ($p<10^{-6}$) but insignificant for the global order $\phi$ (p=0.086).
The difference between the two is likely due to rare fragmentations, which we see in large groups $\gtrsim 100$, 
noting also that $\phi$ is systematically lower than $\phiC$. The fact that there is a 
significant increase in $\phiC$ is perhaps even 
more surprising than the similar effect apparent in 
Fig~\ref{fig:noise}(a3). The magnitude of the increase in order $\phiC$ from $\eta_\theta=0$ to $\eta_\theta\sim 5^\circ$ that 
is apparent in Fig~\ref{fig:noise}(a2) is about 1\% (a relatively large difference: the mis-ordering halves).
However, the corresponding increase in  Fig~\ref{fig:noise}(b2) is nearly an order of magnitude weaker and occurs at much 
smaller noise $\eta\sim 0.5^\circ$. This signifies a different mechanism for the much weaker increase in order that occurs 
under such post-decision noise. We speculate that this might be due to subtle changes in the swarm structure resulting from 
the addition of noise, noting that the density is systematically lower in the presence of weak post-decision noise (see SI for details). 
Such changes could plausibly affect path-entropy maximising trajectories in such a way that 
they generate a higher order. Although there is no obvious intuitive explanation for this it could be related to the 
fact that the agents have more information on the global organisation at lower densities, where there are fewer particle 
overlaps in the visual state.

To conclude, we analyse a simple model that could underly evolutionary fitness and 
hence intelligent behaviour. This model involves agents that seek to maximise the 
path entropy of their future trajectories, analogous to keeping future options open. 
The entropy is here computed over visual states, such as would be perceived by animals 
that rely primarily on vision to sense and navigate the world around them. 
Such path-entropy maximisation strategies could be of broader interest within biology, e.g. 
in the biochemical state space accessible to micro-organisms or cells. However, we believe 
that it will be easier to test these ideas in higher animals that exhibit swarming motion 
where the state space is lower dimensional and the dynamics of inertial flying (or swimming)
 agents is much more simple and well understood than the nonlinear chemical kinetics of 
 cellular biochemistry. 

We find that the ``bottom-up'' principle of maximisation of path entropy is a 
promising candidate to understand the emergence of properties like co-alignment and 
cohesion observed in typical swarming phenotypes. This principle also leads to 
flocks with opacity values close to 0.5, in agreement with observations on some bird flocks \cite{Pearce10422}.

Although the algorithm is highly computationally demanding 
it involves a simple mapping from an observed visual state to an action. 
Heuristics that mimic this process and that could operate under animal
cognition in real time are easy to develop. For example, an artificial neural
network could be trained on simulation data to choose actions from sensory 
input. Similar algorithms could also find use in novel forms of active, 
information-processing (``intelligent'') matter that may soon form part of the experimental
landscape.

\begin{acknowledgments}
Funding was provided by UK Engineering and Physical Sciences Research Council though the Mathematics for Real World Systems Centre for Doctoral Training grant no. EP/L015374/1 (H.L.D.). All numerical work was carried out using the Scientific Computing Research Technology Platform of the University of Warwick. M.S.T. acknowledges the support of a long-term fellowship from the Japan Society for the Promotion of Science, a Leverhulme Trust visiting fellowship and the peerless hospitality of Prof. Ryoichi Yamamoto (Kyoto).
\end{acknowledgments}

\nocite{*}

\newpage
  \section*{S1: Fragmentation dependent on $\tau$}
\begin{figure}[ht!]
      \centering
      \includegraphics[width=\columnwidth]{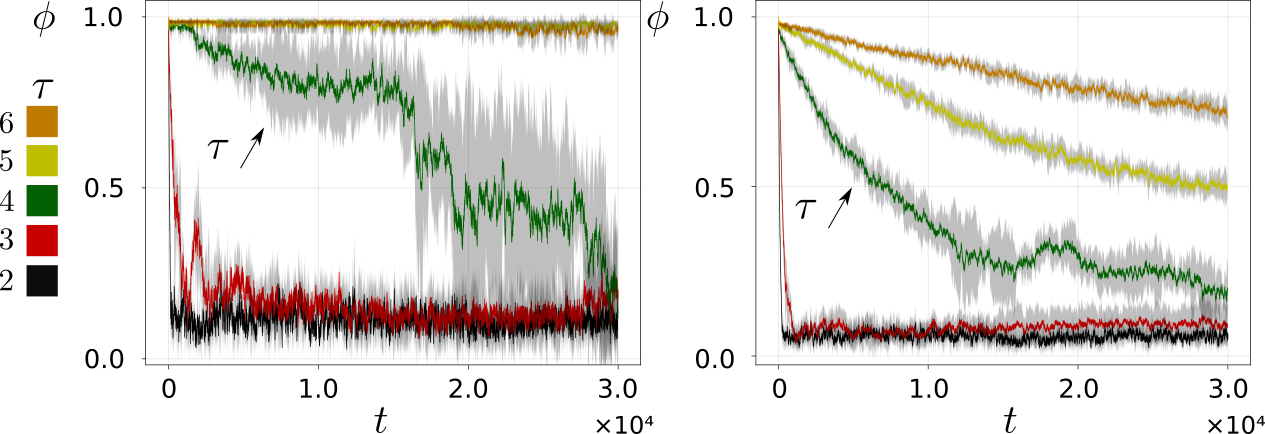}
      \caption{Global order $\phi$ decays over time in large swarms 
      without noise, here $N=250$ (left) is compared with  
      $N=50$ (right), the latter being more stable without noise. 
      The insets show data for longer times $T\sim 10^{4}$. 
     }
      \label{fig:s1}
  \end{figure}

  Over long time scales we sometimes find that there is a small rate of fragmentation, i.e. individuals leave the main group. This rate dramatically decreases with increasing $\tau$ as shown in \ref{fig:s1}.
  \section*{S2: Clustering Methodology}
  Determining clusters from data is a difficult problem to generalise often 
  requiring the tuning of clustering algorithm hyper-parameters. 
  These parameters are either determined by domain knowledge of the data to 
  be clustered, or by a data-driven discovery process. To detect clusters in 
  our data we employ the density based clustering algorithm 
  DBScan \cite{ester1996density} to identify clusters using agent
   position and orientation data.  Here we opt to determine the
    clustering hyper-parameters \textbf{Eps} and \textbf{MinPts} 
    (as named by M. Ester \textit{et al}) by physical argument. 
    Briefly \textbf{Eps} is a parameter which effectively defines a 
    distance cut-off as to whether points are in the same cluster 
    (this is a \textit{transitive} relationship so two points may be
     mutually too far, but connected by an intermediate point), 
     and \textbf{MinPts} is simply a number for the smallest cluster size.
      \textbf{MinPts} is simple for our application, we wish to identify 
      single agents as ``clusters'' so \textbf{MinPts} $=1$.
  
  For \textbf{Eps} we must first define our distance function. 
  To do this at a given time $t$ in a simulation we take the agents' 
  positions $\boldsymbol{x}_{i}(t) = [x_{i}(t),y_{i}(t)]^{T}$ 
  and orientations $\theta_{i}(t)$ (interpreted in the range $[0,2\pi)$) 
  to compute the distance matrix, 
  \begin{align}
      \boldsymbol{d}_{ij}(t) &= \frac{1}{2}(\boldsymbol{d}_{ij}^{\boldsymbol{x}}(t) + \boldsymbol{d}_{ij}^{\theta}(t)), \label{eq:dist} \\
      \boldsymbol{d}_{ij}^{\boldsymbol{x}}(t) &= \biggl(\frac{\sqrt{(x_{i}(t)-x_{j}(t))^{2}+(y_{i}(t)-y_{j}(t))^2}}{\pi\sqrt{N}}\biggr), \label{eq:d-space}\\
      \boldsymbol{d}_{ij}^{\theta}(t) &= \frac{\text{Min}(|\theta_{i}(t)-\theta_{j}(t)|, 2\pi - |\theta_{i}(t)-\theta_{j}(t)|)}{2\Delta \theta}. \label{eq:d-theta}
  \end{align}
  Where distances have been scaled using the average of the Euclidean distance for the positions, scaled to units of the inter-agent distance at marginal opacity $\sim \pi\sqrt{N}$ (equation \ref{eq:d-space}), and the $2\pi$-Periodic Euclidean distance on the orientation data, scaled to units of the orientational move parameter: $\Delta \theta$ (equation \ref{eq:d-theta}).
  The reasoning for this choice is that we wish to identify highly ordered 
  cohesive groups as clusters, the spatial part of our distance covers 
  the cohesive characteristic and the orientational part covers the ordered 
  characteristic. We choose to scale our distances in space and orientation 
  separately and additively combine them (with a normalisation factor) 
  so that we can choose \textbf{Eps} $=1$. 
  The distance scaling is a reasonable choice given our wish to identify 
  clusters as cohesive, ordered groups targeting marginal opacity, 
  and that our algorithm does target marginal opacity. 
  The distance scaling should also generalise well for other 
  collective motion data targeting marginal opacity. 
  The orientation distance scaling is more arbitrary (in general),
   and if we change the $\Delta \theta$ parameter in our model may need to 
   be re-thought, but here $\Delta \theta = 15^{\circ}$ is generally 
   considered a constant, so we consider it a reasonable choice 
   for the data we present. We chose $\Delta \theta$ over $\Delta \theta$. 
   An alternative method could take an order threshold, say $\phi > 0.9$, 
   and use this to compute an approximate angle to use as a scaling 
   from the definition of $\phi$. However this method then falls to a 
   justification of a particular $\phi$-value. 
   For example taking small $\theta_{i}\sim 0$ we could write the order as 
   $\phi \sim \langle \cos{\theta_{i}}\rangle \sim 1-\frac{\langle \theta_{i}^{2}\rangle}{2}$ so $\sqrt{\langle\theta_{i}^{2}\rangle}\sim \sqrt{2(1-\phi)}$, for $\phi=0.9$ this gives approximately $25^{\circ}$. 
  \section*{S3: Ray-tracing the Visual State}
  \begin{figure}[ht!]
      \centering
      \includegraphics[width=\columnwidth]{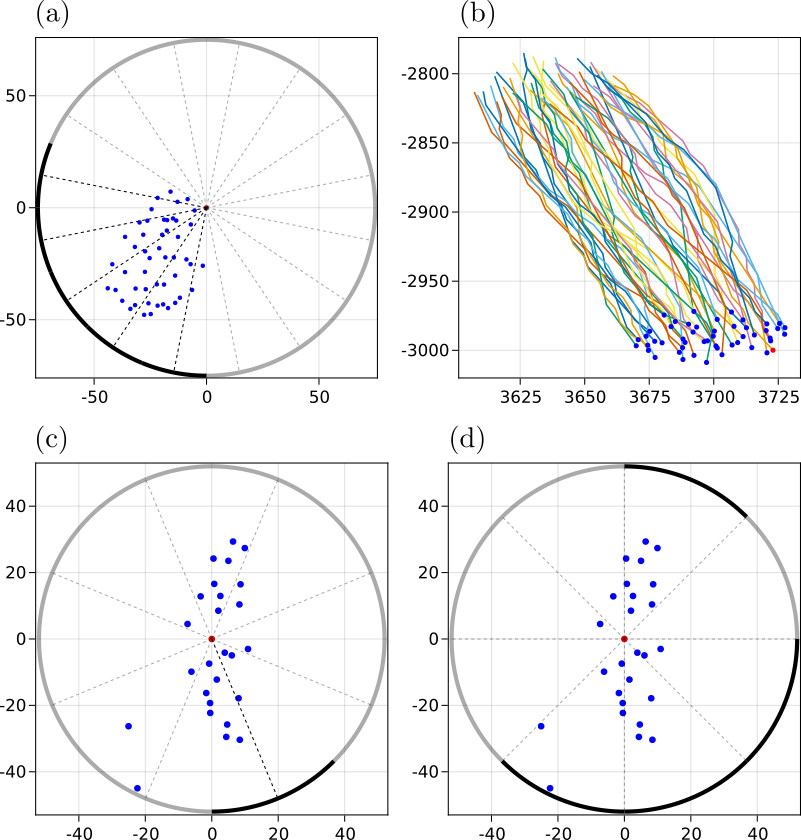}
      \caption{Ray-tracing is an alternative to the 
      visual state representation in the main text. 
      In (a) for each sensor (marked by thick outlined circle) 
      a single ray is projected from an agent's position 
      with an angle such that $0^{\circ}$ points along the 
      agent's current heading direction. 
      If this ray intersects any other agent the entire 
      sensor is considered activated. (b) snapshot of trajectories computed using this method where $N=50,\tau=4,n_{s}=16,\Delta \theta=15^{\circ}, \Delta v = 2$, and $v_{0}=10$.
      (c) and (d) a group using the same parameters with 
      $n_{s}=8$ sensors and the ray-tracing approach. 
      In (c) the visual state is calculated by ray-tracing, 
      in (d) a visual state calculated as in the main text 
      is shown for the situation in (c) for comparison. 
      The dashed lines in (d) indicate the boundaries of 
      sensors, not rays as in (c).
      }
      \label{fig:ray}
  \end{figure}
  An alternative to the integral 
  equation for the visual state is equation \ref{psi_i_n}.
  \begin{equation}
  {\psi}_{i}^{n}=\Theta\Big[\sum_{j}I_{ij}(\chi_{n})I'_{ij}(\chi_{n})\Big] \label{psi_i_n}
  \end{equation}
  That is sensor activation only requires that a line along the 
  unit vector $\hat{\bf n}_{i} = R(\chi_{n})\hat{{\bf v}}_{i}$ 
  intersects one agent $j\neq i$. 
  Here the angles $\chi_{n}=(n-1/2)\frac{2\pi}{n_{s}}$, 
  for $n\in{0,1,2,\ldots,n_{s}}$, are marked with a subscript. 
  The motivation is practical. 
  Computing the visual state integral na\"{i}vely, or even 
  sorting/merging visual projection intervals as we do in the 
  simulations, is computationally intensive. 
  This ray-tracing method requires only the $N-1$ agent visual 
  projections and distance tests already computed for the integral. 
  The draw back is due to the coarser nature of the ray-tracing method.
  
  Using this definition of the visual state does also produce spontaneous order motion. 
  However the results obtained via this ray-tracing approach are not necessarily commensurate with the main text.
  In particular na\"{i}vely applying equation \ref{psi_i_n} with the same number of sensors, $n_{s}$, 
  is clearly a coarse approximation of the visual state as used in the main text. However simply increasing the 
  number of rays will result in a sparse space of possible visual states, the number of possible states scales 
  as $2^{n_{s}}$. In a forthcoming article we will discuss these differences, including methods for aggregating ``sub-rays''.
  That is mathematically one could convolve the visual state space to reduce the effective degrees of freedom 
  from any very large value of $n_s$, e.g. by using $\psi'_n=\sum_{n'=1}^{n_s}G(|n-n'|)\psi_{n'}$ with the kernal $G$
   having some range characteristic of visual resolution. This can be done with a ray-tracing method or the method in the main 
   text, and would represent a ``post-processing'' present in the visual cortex of animals.
  \section*{S4: Density Variation with Noise}
\begin{figure}[ht!]
      \centering
      \includegraphics[width=\columnwidth]{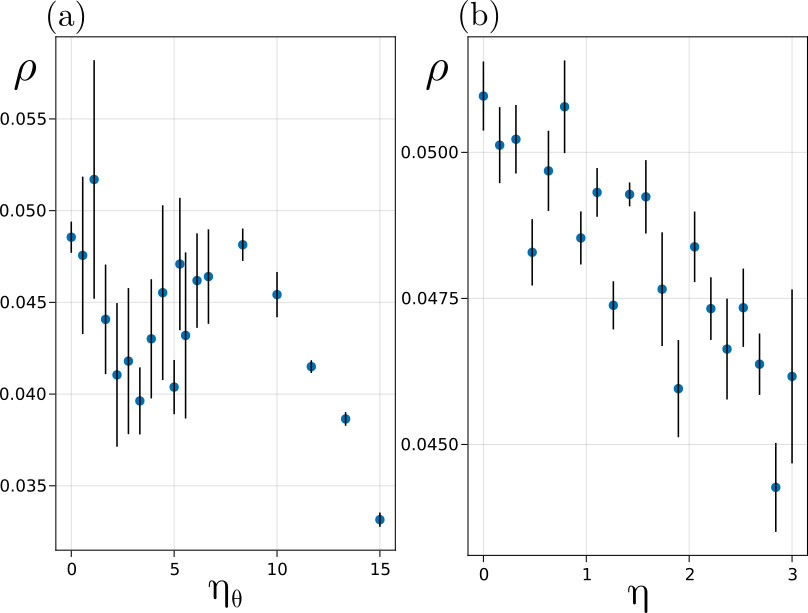}
      \caption{Density variations with (a) cognitive (b) post-decision noise processes for $N=250$ and $\tau=6$. Note the decrease in density in (b) which could account for the counter-intuitive increase in order at small noise values. }
      \label{fig:rho}
  \end{figure}

  Figure \ref{fig:rho} depicts the density for $N=250$ groups simulated using $\tau=6$ under both cognitive and post-decision noise. The variation in density in the latter case could be associated with the increase in order at small noise levels in this case.
  \newpage
  \section*{S5: Velocity Auto-correlation}
\begin{figure}[h]
      \centering
      \includegraphics[width=\columnwidth]{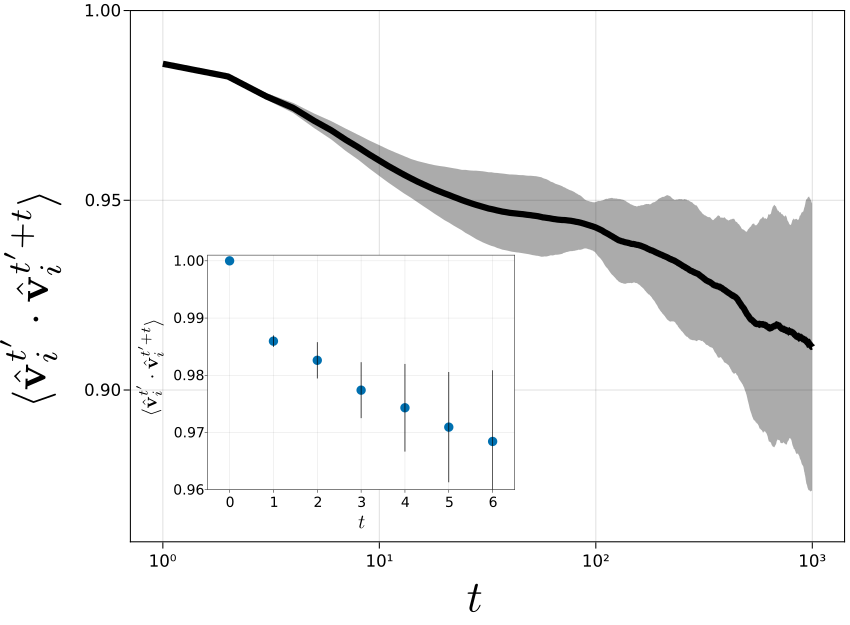}
      \caption{The auto-correlation function of $N=250$ 
      agent velocities with $\tau=6$ and without any added 
      noise shows persistent correlation with a 
      longest correlation time $O(10^5)$. 
      The solid line is the average over 10 realisations 
      with one standard deviation variations in grey. 
      The correlation function was computed using trajectory times $1$ to $1000$, on data simulated up to a longer time $T=2000$. Otherwise statistical artifacts
      appear as e.g from time $t'=1900$ only time lags up to $t=100$ are possible. 
      }
      \label{fig:vdotv}
  \end{figure}

  The velocity autocorrelation function can be used to estimate intrinsic noise levels, as discussed in the main text, see Fig~\ref{fig:vdotv}.
  
  \section*{S6 role of parameter values}
\begin{figure}
      \centering
      \includegraphics[width=\columnwidth]{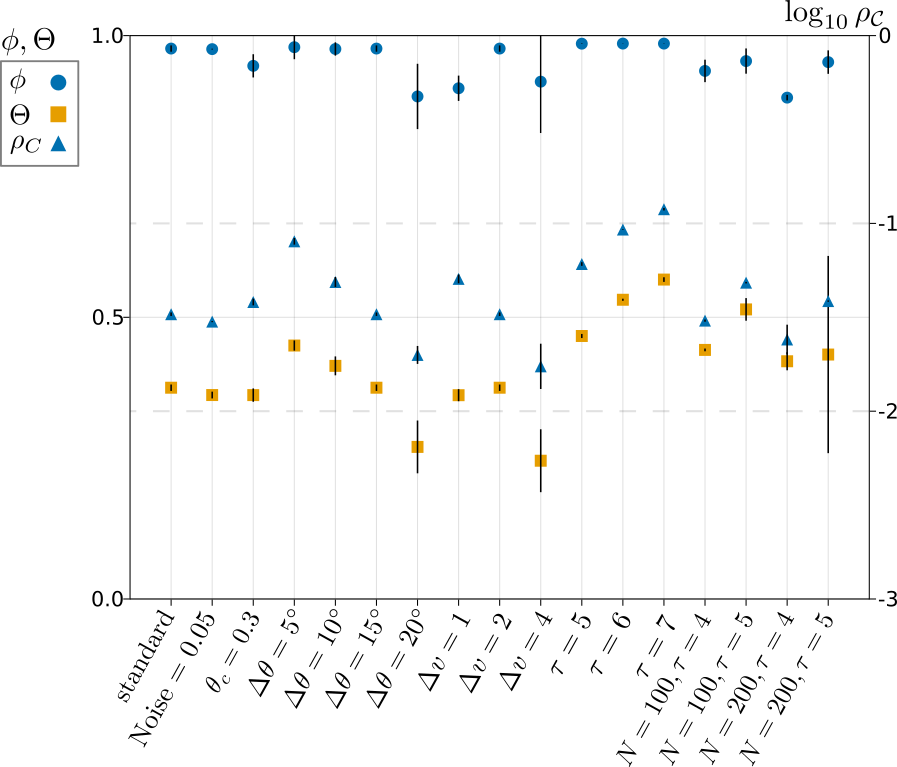}
      \caption{Parameter sweep and the corresponding global order $\phi$, opacity $\Theta$ and largest cluster density $\rho_{\mathcal{C}}$.
      Largest cluster density is used to account for fragmentation occuring in some parameter sets but not others, e.g $\Delta v = 4$.
      Error bars indicate one standard deviation. The standard parameter sets consists of $N=50, \tau=4, n_s=40, \Delta v = 2, v_{0}=10, \Delta \theta = 15^\circ$
      and $\theta_c=0.5$ the threshold of coverage at which sensors are considered filled. The x-axis indicates which parameters
      deviate from the standard set in each case.}
      \label{fig:psweep-no-occ}
  \end{figure}

  In Fig~\ref{fig:psweep-no-occ} we analyse the role of the various control parameters and find that ordered groups with similar density and opacity emerge over a wide regime of control parameters, confirming that fine-tuning is not necessary to obtain such phenotypes.
  
  \section*{S7 SI Movie captions}
  \subsection*{Movie M1}
  Parameters $N=50, \tau=5, n_{s}=40, \Delta v = 2, v_{0}=10, \Delta \theta=15^{\circ}$, with no noise processes.
  \subsection*{Movie M2}
  Parameters $N=50, \tau=5, n_{s}=40, \Delta v = 2, v_{0}=1, \Delta \theta=15^{\circ}$, with no noise processes.
  \subsection*{Movie M3}
  Parameters $N=50, \tau=5, n_{s}=40, \Delta v = 1, v_{0}=2, \Delta \theta=30^{\circ}$, with no noise processes.
  \subsection*{Movie M4}
  Parameters $N=250, \tau=5, n_{s}=40, \Delta v = 2, v_{0}=1, \Delta \theta=30^{\circ}$, with no noise processes.
  \subsection*{Movie M5}
  Parameters $N=250, \tau=6, n_{s}=40, \Delta v = 2, v_{0}=10, \Delta \theta=15^{\circ}$, with cognitive noise parameters $\eta_{v}=0,\eta_{\theta}=5^{\circ}$.
  \subsection*{Movie M6}
  Parameters $N=250, \tau=6, n_{s}=40, \Delta v = 2, v_{0}=10, \Delta \theta=15^{\circ}$, with post-decision noise parameter $\eta=0.474$.
  \subsection*{Movie M7}
  Parameters $N=50, \tau=6, n_{s}=40, \Delta v = 2, v_{0}=10, \Delta \theta=15^{\circ}$, with cognitive noise parameters $\eta_{v}=0,\eta_{\theta}=9.474^{\circ}$.


\begin{thebibliography}{39}%
  \makeatletter
  \providecommand \@ifxundefined [1]{%
   \@ifx{#1\undefined}
  }%
  \providecommand \@ifnum [1]{%
   \ifnum #1\expandafter \@firstoftwo
   \else \expandafter \@secondoftwo
   \fi
  }%
  \providecommand \@ifx [1]{%
   \ifx #1\expandafter \@firstoftwo
   \else \expandafter \@secondoftwo
   \fi
  }%
  \providecommand \natexlab [1]{#1}%
  \providecommand \enquote  [1]{``#1''}%
  \providecommand \bibnamefont  [1]{#1}%
  \providecommand \bibfnamefont [1]{#1}%
  \providecommand \citenamefont [1]{#1}%
  \providecommand \href@noop [0]{\@secondoftwo}%
  \providecommand \href [0]{\begingroup \@sanitize@url \@href}%
  \providecommand \@href[1]{\@@startlink{#1}\@@href}%
  \providecommand \@@href[1]{\endgroup#1\@@endlink}%
  \providecommand \@sanitize@url [0]{\catcode `\\12\catcode `\$12\catcode
    `\&12\catcode `\#12\catcode `\^12\catcode `\_12\catcode `\%12\relax}%
  \providecommand \@@startlink[1]{}%
  \providecommand \@@endlink[0]{}%
  \providecommand \url  [0]{\begingroup\@sanitize@url \@url }%
  \providecommand \@url [1]{\endgroup\@href {#1}{\urlprefix }}%
  \providecommand \urlprefix  [0]{URL }%
  \providecommand \Eprint [0]{\href }%
  \providecommand \doibase [0]{https://doi.org/}%
  \providecommand \selectlanguage [0]{\@gobble}%
  \providecommand \bibinfo  [0]{\@secondoftwo}%
  \providecommand \bibfield  [0]{\@secondoftwo}%
  \providecommand \translation [1]{[#1]}%
  \providecommand \BibitemOpen [0]{}%
  \providecommand \bibitemStop [0]{}%
  \providecommand \bibitemNoStop [0]{.\EOS\space}%
  \providecommand \EOS [0]{\spacefactor3000\relax}%
  \providecommand \BibitemShut  [1]{\csname bibitem#1\endcsname}%
  \let\auto@bib@innerbib\@empty
  \bibitem [{\citenamefont {Sokolov}\ \emph {et~al.}(2007)\citenamefont
    {Sokolov}, \citenamefont {Aranson}, \citenamefont {Kessler},\ and\
    \citenamefont {Goldstein}}]{Sokolov2007Concentration}%
    \BibitemOpen
    \bibfield  {author} {\bibinfo {author} {\bibfnamefont {A.}~\bibnamefont
    {Sokolov}}, \bibinfo {author} {\bibfnamefont {I.~S.}\ \bibnamefont
    {Aranson}}, \bibinfo {author} {\bibfnamefont {J.~O.}\ \bibnamefont
    {Kessler}},\ and\ \bibinfo {author} {\bibfnamefont {R.~E.}\ \bibnamefont
    {Goldstein}},\ }\bibfield  {title} {\bibinfo {title} {Concentration
    dependence of the collective dynamics of swimming bacteria},\ }\href
    {https://doi.org/10.1103/PhysRevLett.98.158102} {\bibfield  {journal}
    {\bibinfo  {journal} {Phys. Rev. Lett.}\ }\textbf {\bibinfo {volume} {98}},\
    \bibinfo {pages} {158102} (\bibinfo {year} {2007})}\BibitemShut {NoStop}%
  \bibitem [{\citenamefont {Vicsek}\ and\ \citenamefont
    {Zafeiris}(2012)}]{vicsek2012collective}%
    \BibitemOpen
    \bibfield  {author} {\bibinfo {author} {\bibfnamefont {T.}~\bibnamefont
    {Vicsek}}\ and\ \bibinfo {author} {\bibfnamefont {A.}~\bibnamefont
    {Zafeiris}},\ }\bibfield  {title} {\bibinfo {title} {Collective motion},\
    }\href@noop {} {\bibfield  {journal} {\bibinfo  {journal} {Physics reports}\
    }\textbf {\bibinfo {volume} {517}},\ \bibinfo {pages} {71} (\bibinfo {year}
    {2012})}\BibitemShut {NoStop}%
  \bibitem [{\citenamefont {Cavagna}\ \emph {et~al.}(2010)\citenamefont
    {Cavagna}, \citenamefont {Cimarelli}, \citenamefont {Giardina}, \citenamefont
    {Parisi}, \citenamefont {Santagati}, \citenamefont {Stefanini},\ and\
    \citenamefont {Viale}}]{Cavagna201005766}%
    \BibitemOpen
    \bibfield  {author} {\bibinfo {author} {\bibfnamefont {A.}~\bibnamefont
    {Cavagna}}, \bibinfo {author} {\bibfnamefont {A.}~\bibnamefont {Cimarelli}},
    \bibinfo {author} {\bibfnamefont {I.}~\bibnamefont {Giardina}}, \bibinfo
    {author} {\bibfnamefont {G.}~\bibnamefont {Parisi}}, \bibinfo {author}
    {\bibfnamefont {R.}~\bibnamefont {Santagati}}, \bibinfo {author}
    {\bibfnamefont {F.}~\bibnamefont {Stefanini}},\ and\ \bibinfo {author}
    {\bibfnamefont {M.}~\bibnamefont {Viale}},\ }\bibfield  {title} {\bibinfo
    {title} {Scale-free correlations in starling flocks},\ }\bibfield  {journal}
    {\bibinfo  {journal} {Proceedings of the National Academy of Sciences}\
    }\href {https://doi.org/10.1073/pnas.1005766107} {10.1073/pnas.1005766107}
    (\bibinfo {year} {2010})\BibitemShut {NoStop}%
  \bibitem [{\citenamefont {Flack}\ \emph {et~al.}(2018)\citenamefont {Flack},
    \citenamefont {Nagy}, \citenamefont {Fiedler}, \citenamefont {Couzin},\ and\
    \citenamefont {Wikelski}}]{flack2018local}%
    \BibitemOpen
    \bibfield  {author} {\bibinfo {author} {\bibfnamefont {A.}~\bibnamefont
    {Flack}}, \bibinfo {author} {\bibfnamefont {M.}~\bibnamefont {Nagy}},
    \bibinfo {author} {\bibfnamefont {W.}~\bibnamefont {Fiedler}}, \bibinfo
    {author} {\bibfnamefont {I.~D.}\ \bibnamefont {Couzin}},\ and\ \bibinfo
    {author} {\bibfnamefont {M.}~\bibnamefont {Wikelski}},\ }\bibfield  {title}
    {\bibinfo {title} {From local collective behavior to global migratory
    patterns in white storks},\ }\href@noop {} {\bibfield  {journal} {\bibinfo
    {journal} {Science}\ }\textbf {\bibinfo {volume} {360}},\ \bibinfo {pages}
    {911} (\bibinfo {year} {2018})}\BibitemShut {NoStop}%
  \bibitem [{\citenamefont {Zitterbart}\ \emph {et~al.}(2011)\citenamefont
    {Zitterbart}, \citenamefont {Wienecke}, \citenamefont {Butler},\ and\
    \citenamefont {Fabry}}]{Zitterbart2011CoordinatedMP}%
    \BibitemOpen
    \bibfield  {author} {\bibinfo {author} {\bibfnamefont {D.~P.}\ \bibnamefont
    {Zitterbart}}, \bibinfo {author} {\bibfnamefont {B.}~\bibnamefont
    {Wienecke}}, \bibinfo {author} {\bibfnamefont {J.~P.}\ \bibnamefont
    {Butler}},\ and\ \bibinfo {author} {\bibfnamefont {B.}~\bibnamefont
    {Fabry}},\ }\bibfield  {title} {\bibinfo {title} {Coordinated movements
    prevent jamming in an emperor penguin huddle},\ }\href
    {https://doi.org/10.1371/journal.pone.0020260} {\bibfield  {journal}
    {\bibinfo  {journal} {PLOS ONE}\ }\textbf {\bibinfo {volume} {6}},\ \bibinfo
    {pages} {1} (\bibinfo {year} {2011})}\BibitemShut {NoStop}%
  \bibitem [{\citenamefont {Moussa{\"\i}d}\ \emph {et~al.}(2011)\citenamefont
    {Moussa{\"\i}d}, \citenamefont {Helbing},\ and\ \citenamefont
    {Theraulaz}}]{Moussa}%
    \BibitemOpen
    \bibfield  {author} {\bibinfo {author} {\bibfnamefont {M.}~\bibnamefont
    {Moussa{\"\i}d}}, \bibinfo {author} {\bibfnamefont {D.}~\bibnamefont
    {Helbing}},\ and\ \bibinfo {author} {\bibfnamefont {G.}~\bibnamefont
    {Theraulaz}},\ }\bibfield  {title} {\bibinfo {title} {How simple rules
    determine pedestrian behavior and crowd disasters},\ }\bibfield  {journal}
    {\bibinfo  {journal} {Proceedings of the National Academy of Sciences}\
    }\href {https://doi.org/10.1073/pnas.1016507108} {10.1073/pnas.1016507108}
    (\bibinfo {year} {2011})\BibitemShut {NoStop}%
  \bibitem [{\citenamefont {Pol}\ \emph {et~al.}(2021)\citenamefont {Pol},
    \citenamefont {Mancuso}, \citenamefont {Smith}, \citenamefont {Marsicano},
    \citenamefont {Ramezani}, \citenamefont {Cerda}, \citenamefont {Otero},\ and\
    \citenamefont {Fernandez}}]{dinosaurs}%
    \BibitemOpen
    \bibfield  {author} {\bibinfo {author} {\bibfnamefont {D.}~\bibnamefont
    {Pol}}, \bibinfo {author} {\bibfnamefont {A.~C.}\ \bibnamefont {Mancuso}},
    \bibinfo {author} {\bibfnamefont {R.~M.~H.}\ \bibnamefont {Smith}}, \bibinfo
    {author} {\bibfnamefont {C.~A.}\ \bibnamefont {Marsicano}}, \bibinfo {author}
    {\bibfnamefont {J.}~\bibnamefont {Ramezani}}, \bibinfo {author}
    {\bibfnamefont {I.~A.}\ \bibnamefont {Cerda}}, \bibinfo {author}
    {\bibfnamefont {A.}~\bibnamefont {Otero}},\ and\ \bibinfo {author}
    {\bibfnamefont {V.}~\bibnamefont {Fernandez}},\ }\bibfield  {title} {\bibinfo
    {title} {Earliest evidence of herd-living and age segregation amongst
    dinosaurs},\ }\href {https://doi.org/10.1038/s41598-021-99176-1} {\bibfield
    {journal} {\bibinfo  {journal} {Scientific Reports}\ }\textbf {\bibinfo
    {volume} {11}},\ \bibinfo {pages} {20023} (\bibinfo {year}
    {2021})}\BibitemShut {NoStop}%
  \bibitem [{\citenamefont {Vicsek}\ \emph {et~al.}(1995)\citenamefont {Vicsek},
    \citenamefont {Czir\'ok}, \citenamefont {Ben-Jacob}, \citenamefont {Cohen},\
    and\ \citenamefont {Shochet}}]{Vicsek1995Novel}%
    \BibitemOpen
    \bibfield  {author} {\bibinfo {author} {\bibfnamefont {T.}~\bibnamefont
    {Vicsek}}, \bibinfo {author} {\bibfnamefont {A.}~\bibnamefont {Czir\'ok}},
    \bibinfo {author} {\bibfnamefont {E.}~\bibnamefont {Ben-Jacob}}, \bibinfo
    {author} {\bibfnamefont {I.}~\bibnamefont {Cohen}},\ and\ \bibinfo {author}
    {\bibfnamefont {O.}~\bibnamefont {Shochet}},\ }\bibfield  {title} {\bibinfo
    {title} {Novel type of phase transition in a system of self-driven
    particles},\ }\href {https://doi.org/10.1103/PhysRevLett.75.1226} {\bibfield
    {journal} {\bibinfo  {journal} {Phys. Rev. Lett.}\ }\textbf {\bibinfo
    {volume} {75}},\ \bibinfo {pages} {1226} (\bibinfo {year}
    {1995})}\BibitemShut {NoStop}%
  \bibitem [{\citenamefont {Buhl}\ \emph {et~al.}(2006)\citenamefont {Buhl},
    \citenamefont {Sumpter}, \citenamefont {Couzin}, \citenamefont {Hale},
    \citenamefont {Despland}, \citenamefont {Miller},\ and\ \citenamefont
    {Simpson}}]{Buhl2006Recent}%
    \BibitemOpen
    \bibfield  {author} {\bibinfo {author} {\bibfnamefont {J.}~\bibnamefont
    {Buhl}}, \bibinfo {author} {\bibfnamefont {D.~J.~T.}\ \bibnamefont
    {Sumpter}}, \bibinfo {author} {\bibfnamefont {I.~D.}\ \bibnamefont {Couzin}},
    \bibinfo {author} {\bibfnamefont {J.~J.}\ \bibnamefont {Hale}}, \bibinfo
    {author} {\bibfnamefont {E.}~\bibnamefont {Despland}}, \bibinfo {author}
    {\bibfnamefont {E.~R.}\ \bibnamefont {Miller}},\ and\ \bibinfo {author}
    {\bibfnamefont {S.~J.}\ \bibnamefont {Simpson}},\ }\bibfield  {title}
    {\bibinfo {title} {From disorder to order in marching locusts},\ }\href
    {https://doi.org/10.1126/science.1125142} {\bibfield  {journal} {\bibinfo
    {journal} {Science}\ }\textbf {\bibinfo {volume} {312}},\ \bibinfo {pages}
    {1402} (\bibinfo {year} {2006})}\BibitemShut {NoStop}%
  \bibitem [{\citenamefont {Szab\'o}\ \emph {et~al.}(2006)\citenamefont
    {Szab\'o}, \citenamefont {Sz\"oll\"osi}, \citenamefont {G\"onci},
    \citenamefont {Jur\'anyi}, \citenamefont {Selmeczi},\ and\ \citenamefont
    {Vicsek}}]{Szabo2006Phase}%
    \BibitemOpen
    \bibfield  {author} {\bibinfo {author} {\bibfnamefont {B.}~\bibnamefont
    {Szab\'o}}, \bibinfo {author} {\bibfnamefont {G.~J.}\ \bibnamefont
    {Sz\"oll\"osi}}, \bibinfo {author} {\bibfnamefont {B.}~\bibnamefont
    {G\"onci}}, \bibinfo {author} {\bibfnamefont {Z.}~\bibnamefont {Jur\'anyi}},
    \bibinfo {author} {\bibfnamefont {D.}~\bibnamefont {Selmeczi}},\ and\
    \bibinfo {author} {\bibfnamefont {T.}~\bibnamefont {Vicsek}},\ }\bibfield
    {title} {\bibinfo {title} {Phase transition in the collective migration of
    tissue cells: Experiment and model},\ }\href
    {https://doi.org/10.1103/PhysRevE.74.061908} {\bibfield  {journal} {\bibinfo
    {journal} {Phys. Rev. E}\ }\textbf {\bibinfo {volume} {74}},\ \bibinfo
    {pages} {061908} (\bibinfo {year} {2006})}\BibitemShut {NoStop}%
  \bibitem [{\citenamefont {Gr{\'{e}}goire}\ and\ \citenamefont
    {Chat{\'{e}}}(2004)}]{Gregoire2004}%
    \BibitemOpen
    \bibfield  {author} {\bibinfo {author} {\bibfnamefont {G.}~\bibnamefont
    {Gr{\'{e}}goire}}\ and\ \bibinfo {author} {\bibfnamefont {H.}~\bibnamefont
    {Chat{\'{e}}}},\ }\bibfield  {title} {\bibinfo {title} {{Onset of Collective
    and Cohesive Motion}},\ }\href
    {https://doi.org/10.1103/PhysRevLett.92.025702} {\bibfield  {journal}
    {\bibinfo  {journal} {Physical Review Letters}\ }\textbf {\bibinfo {volume}
    {92}},\ \bibinfo {pages} {025702} (\bibinfo {year} {2004})}\BibitemShut
    {NoStop}%
  \bibitem [{\citenamefont {Chat{\'{e}}}\ \emph {et~al.}(2008)\citenamefont
    {Chat{\'{e}}}, \citenamefont {Ginelli}, \citenamefont {Gr{\'{e}}goire},
    \citenamefont {Peruani},\ and\ \citenamefont {Raynaud}}]{Chate2008}%
    \BibitemOpen
    \bibfield  {author} {\bibinfo {author} {\bibfnamefont {H.}~\bibnamefont
    {Chat{\'{e}}}}, \bibinfo {author} {\bibfnamefont {F.}~\bibnamefont
    {Ginelli}}, \bibinfo {author} {\bibfnamefont {G.}~\bibnamefont
    {Gr{\'{e}}goire}}, \bibinfo {author} {\bibfnamefont {F.}~\bibnamefont
    {Peruani}},\ and\ \bibinfo {author} {\bibfnamefont {F.}~\bibnamefont
    {Raynaud}},\ }\bibfield  {title} {\bibinfo {title} {{Modeling collective
    motion: Variations on the Vicsek model}},\ }\href
    {https://doi.org/10.1140/epjb/e2008-00275-9} {\bibfield  {journal} {\bibinfo
    {journal} {European Physical Journal B}\ }\textbf {\bibinfo {volume} {64}},\
    \bibinfo {pages} {451} (\bibinfo {year} {2008})}\BibitemShut {NoStop}%
  \bibitem [{\citenamefont {Ginelli}\ and\ \citenamefont
    {Chat\'e}(2010)}]{PhysRevLettMetricFree}%
    \BibitemOpen
    \bibfield  {author} {\bibinfo {author} {\bibfnamefont {F.}~\bibnamefont
    {Ginelli}}\ and\ \bibinfo {author} {\bibfnamefont {H.}~\bibnamefont
    {Chat\'e}},\ }\bibfield  {title} {\bibinfo {title} {Relevance of metric-free
    interactions in flocking phenomena},\ }\href
    {https://doi.org/10.1103/PhysRevLett.105.168103} {\bibfield  {journal}
    {\bibinfo  {journal} {Phys. Rev. Lett.}\ }\textbf {\bibinfo {volume} {105}},\
    \bibinfo {pages} {168103} (\bibinfo {year} {2010})}\BibitemShut {NoStop}%
  \bibitem [{\citenamefont {Weber}\ \emph {et~al.}(2013)\citenamefont {Weber},
    \citenamefont {Hanke}, \citenamefont {Deseigne}, \citenamefont {L\'eonard},
    \citenamefont {Dauchot}, \citenamefont {Frey},\ and\ \citenamefont
    {Chat\'e}}]{Weber2013Long}%
    \BibitemOpen
    \bibfield  {author} {\bibinfo {author} {\bibfnamefont {C.~A.}\ \bibnamefont
    {Weber}}, \bibinfo {author} {\bibfnamefont {T.}~\bibnamefont {Hanke}},
    \bibinfo {author} {\bibfnamefont {J.}~\bibnamefont {Deseigne}}, \bibinfo
    {author} {\bibfnamefont {S.}~\bibnamefont {L\'eonard}}, \bibinfo {author}
    {\bibfnamefont {O.}~\bibnamefont {Dauchot}}, \bibinfo {author} {\bibfnamefont
    {E.}~\bibnamefont {Frey}},\ and\ \bibinfo {author} {\bibfnamefont
    {H.}~\bibnamefont {Chat\'e}},\ }\bibfield  {title} {\bibinfo {title}
    {Long-range ordering of vibrated polar disks},\ }\href
    {https://doi.org/10.1103/PhysRevLett.110.208001} {\bibfield  {journal}
    {\bibinfo  {journal} {Phys. Rev. Lett.}\ }\textbf {\bibinfo {volume} {110}},\
    \bibinfo {pages} {208001} (\bibinfo {year} {2013})}\BibitemShut {NoStop}%
  \bibitem [{\citenamefont {Calovi}\ \emph {et~al.}(2014)\citenamefont {Calovi},
    \citenamefont {Lopez}, \citenamefont {Ngo}, \citenamefont {Sire},
    \citenamefont {Chat'e},\ and\ \citenamefont
    {Theraulaz}}]{Calovi2014SwarmingSM}%
    \BibitemOpen
    \bibfield  {author} {\bibinfo {author} {\bibfnamefont {D.~S.}\ \bibnamefont
    {Calovi}}, \bibinfo {author} {\bibfnamefont {U.}~\bibnamefont {Lopez}},
    \bibinfo {author} {\bibfnamefont {S.}~\bibnamefont {Ngo}}, \bibinfo {author}
    {\bibfnamefont {C.}~\bibnamefont {Sire}}, \bibinfo {author} {\bibfnamefont
    {H.}~\bibnamefont {Chat'e}},\ and\ \bibinfo {author} {\bibfnamefont
    {G.}~\bibnamefont {Theraulaz}},\ }\bibfield  {title} {\bibinfo {title}
    {Swarming, schooling, milling: phase diagram of a data-driven fish school
    model},\ }\href@noop {} {\bibfield  {journal} {\bibinfo  {journal} {New
    Journal of Physics}\ }\textbf {\bibinfo {volume} {16}},\ \bibinfo {pages}
    {015026} (\bibinfo {year} {2014})}\BibitemShut {NoStop}%
  \bibitem [{\citenamefont {Nishiguchi}\ \emph {et~al.}(2017)\citenamefont
    {Nishiguchi}, \citenamefont {Nagai}, \citenamefont {Chat\'e},\ and\
    \citenamefont {Sano}}]{Nishiguchi2017Long}%
    \BibitemOpen
    \bibfield  {author} {\bibinfo {author} {\bibfnamefont {D.}~\bibnamefont
    {Nishiguchi}}, \bibinfo {author} {\bibfnamefont {K.~H.}\ \bibnamefont
    {Nagai}}, \bibinfo {author} {\bibfnamefont {H.}~\bibnamefont {Chat\'e}},\
    and\ \bibinfo {author} {\bibfnamefont {M.}~\bibnamefont {Sano}},\ }\bibfield
    {title} {\bibinfo {title} {Long-range nematic order and anomalous
    fluctuations in suspensions of swimming filamentous bacteria},\ }\href
    {https://doi.org/10.1103/PhysRevE.95.020601} {\bibfield  {journal} {\bibinfo
    {journal} {Phys. Rev. E}\ }\textbf {\bibinfo {volume} {95}},\ \bibinfo
    {pages} {020601} (\bibinfo {year} {2017})}\BibitemShut {NoStop}%
  \bibitem [{\citenamefont {Ballerini}\ \emph
    {et~al.}(2008{\natexlab{a}})\citenamefont {Ballerini}, \citenamefont
    {Cabibbo}, \citenamefont {Candelier}, \citenamefont {Cavagna}, \citenamefont
    {Cisbani}, \citenamefont {Giardina}, \citenamefont {Lecomte}, \citenamefont
    {Orlandi}, \citenamefont {Parisi}, \citenamefont {Procaccini}, \citenamefont
    {Viale},\ and\ \citenamefont {Zdravkovic}}]{Ballerini1232Topological}%
    \BibitemOpen
    \bibfield  {author} {\bibinfo {author} {\bibfnamefont {M.}~\bibnamefont
    {Ballerini}}, \bibinfo {author} {\bibfnamefont {N.}~\bibnamefont {Cabibbo}},
    \bibinfo {author} {\bibfnamefont {R.}~\bibnamefont {Candelier}}, \bibinfo
    {author} {\bibfnamefont {A.}~\bibnamefont {Cavagna}}, \bibinfo {author}
    {\bibfnamefont {E.}~\bibnamefont {Cisbani}}, \bibinfo {author} {\bibfnamefont
    {I.}~\bibnamefont {Giardina}}, \bibinfo {author} {\bibfnamefont
    {V.}~\bibnamefont {Lecomte}}, \bibinfo {author} {\bibfnamefont
    {A.}~\bibnamefont {Orlandi}}, \bibinfo {author} {\bibfnamefont
    {G.}~\bibnamefont {Parisi}}, \bibinfo {author} {\bibfnamefont
    {A.}~\bibnamefont {Procaccini}}, \bibinfo {author} {\bibfnamefont
    {M.}~\bibnamefont {Viale}},\ and\ \bibinfo {author} {\bibfnamefont
    {V.}~\bibnamefont {Zdravkovic}},\ }\bibfield  {title} {\bibinfo {title}
    {Interaction ruling animal collective behavior depends on topological rather
    than metric distance: Evidence from a field study},\ }\href
    {https://doi.org/10.1073/pnas.0711437105} {\bibfield  {journal} {\bibinfo
    {journal} {Proceedings of the National Academy of Sciences}\ }\textbf
    {\bibinfo {volume} {105}},\ \bibinfo {pages} {1232} (\bibinfo {year}
    {2008}{\natexlab{a}})}\BibitemShut {NoStop}%
  \bibitem [{\citenamefont {Hildenbrandt}\ \emph {et~al.}(2010)\citenamefont
    {Hildenbrandt}, \citenamefont {Carere},\ and\ \citenamefont
    {Hemelrijk}}]{starlingdisplays}%
    \BibitemOpen
    \bibfield  {author} {\bibinfo {author} {\bibfnamefont {H.}~\bibnamefont
    {Hildenbrandt}}, \bibinfo {author} {\bibfnamefont {C.}~\bibnamefont
    {Carere}},\ and\ \bibinfo {author} {\bibfnamefont {C.}~\bibnamefont
    {Hemelrijk}},\ }\bibfield  {title} {\bibinfo {title} {{Self-organized aerial
    displays of thousands of starlings: a model}},\ }\href
    {https://doi.org/10.1093/beheco/arq149} {\bibfield  {journal} {\bibinfo
    {journal} {Behavioral Ecology}\ }\textbf {\bibinfo {volume} {21}},\ \bibinfo
    {pages} {1349} (\bibinfo {year} {2010})}\BibitemShut {NoStop}%
  \bibitem [{\citenamefont {Pearce}\ \emph {et~al.}(2014)\citenamefont {Pearce},
    \citenamefont {Miller}, \citenamefont {Rowlands},\ and\ \citenamefont
    {Turner}}]{Pearce10422}%
    \BibitemOpen
    \bibfield  {author} {\bibinfo {author} {\bibfnamefont {D.~J.~G.}\
    \bibnamefont {Pearce}}, \bibinfo {author} {\bibfnamefont {A.~M.}\
    \bibnamefont {Miller}}, \bibinfo {author} {\bibfnamefont {G.}~\bibnamefont
    {Rowlands}},\ and\ \bibinfo {author} {\bibfnamefont {M.~S.}\ \bibnamefont
    {Turner}},\ }\bibfield  {title} {\bibinfo {title} {Role of projection in the
    control of bird flocks},\ }\href {https://doi.org/10.1073/pnas.1402202111}
    {\bibfield  {journal} {\bibinfo  {journal} {Proceedings of the National
    Academy of Sciences}\ }\textbf {\bibinfo {volume} {111}},\ \bibinfo {pages}
    {10422} (\bibinfo {year} {2014})}\BibitemShut {NoStop}%
  \bibitem [{\citenamefont {Bastien}\ and\ \citenamefont
    {Romanczuk}(2020)}]{Pawel2020}%
    \BibitemOpen
    \bibfield  {author} {\bibinfo {author} {\bibfnamefont {R.}~\bibnamefont
    {Bastien}}\ and\ \bibinfo {author} {\bibfnamefont {P.}~\bibnamefont
    {Romanczuk}},\ }\bibfield  {title} {\bibinfo {title} {A model of collective
    behavior based purely on vision},\ }\href
    {https://doi.org/10.1126/sciadv.aay0792} {\bibfield  {journal} {\bibinfo
    {journal} {Science Advances}\ }\textbf {\bibinfo {volume} {6}},\ \bibinfo
    {pages} {eaay0792} (\bibinfo {year} {2020})},\ \Eprint
    {https://arxiv.org/abs/https://www.science.org/doi/pdf/10.1126/sciadv.aay0792}
    {https://www.science.org/doi/pdf/10.1126/sciadv.aay0792} \BibitemShut
    {NoStop}%
  \bibitem [{\citenamefont {Gallup}\ \emph {et~al.}(2012)\citenamefont {Gallup},
    \citenamefont {Hale}, \citenamefont {Sumpter}, \citenamefont {Garnier},
    \citenamefont {Kacelnik}, \citenamefont {Krebs},\ and\ \citenamefont
    {Couzin}}]{Gallup7245humanvisinfo}%
    \BibitemOpen
    \bibfield  {author} {\bibinfo {author} {\bibfnamefont {A.~C.}\ \bibnamefont
    {Gallup}}, \bibinfo {author} {\bibfnamefont {J.~J.}\ \bibnamefont {Hale}},
    \bibinfo {author} {\bibfnamefont {D.~J.~T.}\ \bibnamefont {Sumpter}},
    \bibinfo {author} {\bibfnamefont {S.}~\bibnamefont {Garnier}}, \bibinfo
    {author} {\bibfnamefont {A.}~\bibnamefont {Kacelnik}}, \bibinfo {author}
    {\bibfnamefont {J.~R.}\ \bibnamefont {Krebs}},\ and\ \bibinfo {author}
    {\bibfnamefont {I.~D.}\ \bibnamefont {Couzin}},\ }\bibfield  {title}
    {\bibinfo {title} {Visual attention and the acquisition of information in
    human crowds},\ }\href {https://doi.org/10.1073/pnas.1116141109} {\bibfield
    {journal} {\bibinfo  {journal} {Proceedings of the National Academy of
    Sciences}\ }\textbf {\bibinfo {volume} {109}},\ \bibinfo {pages} {7245}
    (\bibinfo {year} {2012})}\BibitemShut {NoStop}%
  \bibitem [{\citenamefont {Attanasi}\ \emph {et~al.}(2014)\citenamefont
    {Attanasi}, \citenamefont {Cavagna}, \citenamefont {Del~Castello},
    \citenamefont {Giardina}, \citenamefont {Grigera}, \citenamefont {Jeli{\'c}},
    \citenamefont {Melillo}, \citenamefont {Parisi}, \citenamefont {Pohl},
    \citenamefont {Shen} \emph {et~al.}}]{attanasi2014Information}%
    \BibitemOpen
    \bibfield  {author} {\bibinfo {author} {\bibfnamefont {A.}~\bibnamefont
    {Attanasi}}, \bibinfo {author} {\bibfnamefont {A.}~\bibnamefont {Cavagna}},
    \bibinfo {author} {\bibfnamefont {L.}~\bibnamefont {Del~Castello}}, \bibinfo
    {author} {\bibfnamefont {I.}~\bibnamefont {Giardina}}, \bibinfo {author}
    {\bibfnamefont {T.~S.}\ \bibnamefont {Grigera}}, \bibinfo {author}
    {\bibfnamefont {A.}~\bibnamefont {Jeli{\'c}}}, \bibinfo {author}
    {\bibfnamefont {S.}~\bibnamefont {Melillo}}, \bibinfo {author} {\bibfnamefont
    {L.}~\bibnamefont {Parisi}}, \bibinfo {author} {\bibfnamefont
    {O.}~\bibnamefont {Pohl}}, \bibinfo {author} {\bibfnamefont {E.}~\bibnamefont
    {Shen}}, \emph {et~al.},\ }\bibfield  {title} {\bibinfo {title} {Information
    transfer and behavioural inertia in starling flocks},\ }\href@noop {}
    {\bibfield  {journal} {\bibinfo  {journal} {Nature physics}\ }\textbf
    {\bibinfo {volume} {10}},\ \bibinfo {pages} {691} (\bibinfo {year}
    {2014})}\BibitemShut {NoStop}%
  \bibitem [{\citenamefont {Toner}\ and\ \citenamefont
    {Tu}(1995)}]{Toner1995Long}%
    \BibitemOpen
    \bibfield  {author} {\bibinfo {author} {\bibfnamefont {J.}~\bibnamefont
    {Toner}}\ and\ \bibinfo {author} {\bibfnamefont {Y.}~\bibnamefont {Tu}},\
    }\bibfield  {title} {\bibinfo {title} {Long-range order in a two-dimensional
    dynamical $\mathrm{XY}$ model: How birds fly together},\ }\href
    {https://doi.org/10.1103/PhysRevLett.75.4326} {\bibfield  {journal} {\bibinfo
     {journal} {Phys. Rev. Lett.}\ }\textbf {\bibinfo {volume} {75}},\ \bibinfo
    {pages} {4326} (\bibinfo {year} {1995})}\BibitemShut {NoStop}%
  \bibitem [{\citenamefont {Durve}\ \emph {et~al.}(2020)\citenamefont {Durve},
    \citenamefont {Peruani},\ and\ \citenamefont {Celani}}]{durve}%
    \BibitemOpen
    \bibfield  {author} {\bibinfo {author} {\bibfnamefont {M.}~\bibnamefont
    {Durve}}, \bibinfo {author} {\bibfnamefont {F.}~\bibnamefont {Peruani}},\
    and\ \bibinfo {author} {\bibfnamefont {A.}~\bibnamefont {Celani}},\
    }\bibfield  {title} {\bibinfo {title} {Learning to flock through
    reinforcement},\ }\href {https://doi.org/10.1103/PhysRevE.102.012601}
    {\bibfield  {journal} {\bibinfo  {journal} {Phys. Rev. E}\ }\textbf {\bibinfo
    {volume} {102}},\ \bibinfo {pages} {012601} (\bibinfo {year}
    {2020})}\BibitemShut {NoStop}%
  \bibitem [{\citenamefont {Nemoto}\ \emph {et~al.}(2019)\citenamefont {Nemoto},
    \citenamefont {Fodor}, \citenamefont {Cates}, \citenamefont {Jack},\ and\
    \citenamefont {Tailleur}}]{j1}%
    \BibitemOpen
    \bibfield  {author} {\bibinfo {author} {\bibfnamefont {T.}~\bibnamefont
    {Nemoto}}, \bibinfo {author} {\bibfnamefont {E.}~\bibnamefont {Fodor}},
    \bibinfo {author} {\bibfnamefont {M.~E.}\ \bibnamefont {Cates}}, \bibinfo
    {author} {\bibfnamefont {R.~L.}\ \bibnamefont {Jack}},\ and\ \bibinfo
    {author} {\bibfnamefont {J.}~\bibnamefont {Tailleur}},\ }\bibfield  {title}
    {\bibinfo {title} {Optimizing active work: Dynamical phase transitions,
    collective motion, and jamming},\ }\href
    {https://doi.org/10.1103/PhysRevE.99.022605} {\bibfield  {journal} {\bibinfo
    {journal} {Phys. Rev. E}\ }\textbf {\bibinfo {volume} {99}},\ \bibinfo
    {pages} {022605} (\bibinfo {year} {2019})}\BibitemShut {NoStop}%
  \bibitem [{\citenamefont {Étienne Fodor}\ \emph {et~al.}(2020)\citenamefont
    {Étienne Fodor}, \citenamefont {Nemoto},\ and\ \citenamefont
    {Vaikuntanathan}}]{j2}%
    \BibitemOpen
    \bibfield  {author} {\bibinfo {author} {\bibnamefont {Étienne Fodor}},
    \bibinfo {author} {\bibfnamefont {T.}~\bibnamefont {Nemoto}},\ and\ \bibinfo
    {author} {\bibfnamefont {S.}~\bibnamefont {Vaikuntanathan}},\ }\bibfield
    {title} {\bibinfo {title} {Dissipation controls transport and phase
    transitions in active fluids: mobility, diffusion and biased ensembles},\
    }\href {https://doi.org/10.1088/1367-2630/ab6353} {\bibfield  {journal}
    {\bibinfo  {journal} {New Journal of Physics}\ }\textbf {\bibinfo {volume}
    {22}},\ \bibinfo {pages} {013052} (\bibinfo {year} {2020})}\BibitemShut
    {NoStop}%
  \bibitem [{\citenamefont {Keta}\ \emph {et~al.}(2021)\citenamefont {Keta},
    \citenamefont {Fodor}, \citenamefont {van Wijland}, \citenamefont {Cates},\
    and\ \citenamefont {Jack}}]{j3}%
    \BibitemOpen
    \bibfield  {author} {\bibinfo {author} {\bibfnamefont {Y.-E.}\ \bibnamefont
    {Keta}}, \bibinfo {author} {\bibfnamefont {E.}~\bibnamefont {Fodor}},
    \bibinfo {author} {\bibfnamefont {F.}~\bibnamefont {van Wijland}}, \bibinfo
    {author} {\bibfnamefont {M.~E.}\ \bibnamefont {Cates}},\ and\ \bibinfo
    {author} {\bibfnamefont {R.~L.}\ \bibnamefont {Jack}},\ }\bibfield  {title}
    {\bibinfo {title} {Collective motion in large deviations of active
    particles},\ }\href {https://doi.org/10.1103/PhysRevE.103.022603} {\bibfield
    {journal} {\bibinfo  {journal} {Phys. Rev. E}\ }\textbf {\bibinfo {volume}
    {103}},\ \bibinfo {pages} {022603} (\bibinfo {year} {2021})}\BibitemShut
    {NoStop}%
  \bibitem [{\citenamefont {Fodor}\ \emph {et~al.}(2022)\citenamefont {Fodor},
    \citenamefont {Jack},\ and\ \citenamefont {Cates}}]{j4}%
    \BibitemOpen
    \bibfield  {author} {\bibinfo {author} {\bibfnamefont {E.}~\bibnamefont
    {Fodor}}, \bibinfo {author} {\bibfnamefont {R.~L.}\ \bibnamefont {Jack}},\
    and\ \bibinfo {author} {\bibfnamefont {M.~E.}\ \bibnamefont {Cates}},\
    }\bibfield  {title} {\bibinfo {title} {Irreversibility and biased ensembles
    in active matter: Insights from stochastic thermodynamics},\ }\href
    {https://doi.org/10.1146/annurev-conmatphys-031720-032419} {\bibfield
    {journal} {\bibinfo  {journal} {Annual Review of Condensed Matter Physics}\
    }\textbf {\bibinfo {volume} {13}},\ \bibinfo {pages} {215} (\bibinfo {year}
    {2022})},\ \Eprint
    {https://arxiv.org/abs/https://doi.org/10.1146/annurev-conmatphys-031720-032419}
    {https://doi.org/10.1146/annurev-conmatphys-031720-032419} \BibitemShut
    {NoStop}%
  \bibitem [{\citenamefont {Wissner-Gross}\ and\ \citenamefont
    {Freer}(2013)}]{wissner2013causal}%
    \BibitemOpen
    \bibfield  {author} {\bibinfo {author} {\bibfnamefont {A.~D.}\ \bibnamefont
    {Wissner-Gross}}\ and\ \bibinfo {author} {\bibfnamefont {C.~E.}\ \bibnamefont
    {Freer}},\ }\bibfield  {title} {\bibinfo {title} {Causal entropic forces},\
    }\href@noop {} {\bibfield  {journal} {\bibinfo  {journal} {Physical review
    letters}\ }\textbf {\bibinfo {volume} {110}},\ \bibinfo {pages} {168702}
    (\bibinfo {year} {2013})}\BibitemShut {NoStop}%
  \bibitem [{\citenamefont {Mohamed}\ and\ \citenamefont
    {Jimenez~Rezende}(2015)}]{empowermentRL}%
    \BibitemOpen
    \bibfield  {author} {\bibinfo {author} {\bibfnamefont {S.}~\bibnamefont
    {Mohamed}}\ and\ \bibinfo {author} {\bibfnamefont {D.}~\bibnamefont
    {Jimenez~Rezende}},\ }\bibfield  {title} {\bibinfo {title} {Variational
    information maximisation for intrinsically motivated reinforcement
    learning},\ }in\ \href
    {https://proceedings.neurips.cc/paper_files/paper/2015/file/e00406144c1e7e35240afed70f34166a-Paper.pdf}
    {\emph {\bibinfo {booktitle} {Advances in Neural Information Processing
    Systems}}},\ Vol.~\bibinfo {volume} {28},\ \bibinfo {editor} {edited by\
    \bibinfo {editor} {\bibfnamefont {C.}~\bibnamefont {Cortes}}, \bibinfo
    {editor} {\bibfnamefont {N.}~\bibnamefont {Lawrence}}, \bibinfo {editor}
    {\bibfnamefont {D.}~\bibnamefont {Lee}}, \bibinfo {editor} {\bibfnamefont
    {M.}~\bibnamefont {Sugiyama}},\ and\ \bibinfo {editor} {\bibfnamefont
    {R.}~\bibnamefont {Garnett}}}\ (\bibinfo  {publisher} {Curran Associates,
    Inc.},\ \bibinfo {year} {2015})\BibitemShut {NoStop}%
  \bibitem [{\citenamefont {Mann}\ and\ \citenamefont
    {Garnett}(2015)}]{Mann20150037}%
    \BibitemOpen
    \bibfield  {author} {\bibinfo {author} {\bibfnamefont {R.~P.}\ \bibnamefont
    {Mann}}\ and\ \bibinfo {author} {\bibfnamefont {R.}~\bibnamefont {Garnett}},\
    }\bibfield  {title} {\bibinfo {title} {The entropic basis of collective
    behaviour},\ }\href {https://doi.org/10.1098/rsif.2015.0037} {\bibfield
    {journal} {\bibinfo  {journal} {Journal of The Royal Society Interface}\
    }\textbf {\bibinfo {volume} {12}},\ \bibinfo {pages} {20150037} (\bibinfo
    {year} {2015})}\BibitemShut {NoStop}%
  \bibitem [{\citenamefont {Charlesworth}\ and\ \citenamefont
    {Turner}(2019)}]{Charlesworth15362}%
    \BibitemOpen
    \bibfield  {author} {\bibinfo {author} {\bibfnamefont {H.~J.}\ \bibnamefont
    {Charlesworth}}\ and\ \bibinfo {author} {\bibfnamefont {M.~S.}\ \bibnamefont
    {Turner}},\ }\bibfield  {title} {\bibinfo {title} {Intrinsically motivated
    collective motion},\ }\href {https://doi.org/10.1073/pnas.1822069116}
    {\bibfield  {journal} {\bibinfo  {journal} {Proceedings of the National
    Academy of Sciences}\ }\textbf {\bibinfo {volume} {116}},\ \bibinfo {pages}
    {15362} (\bibinfo {year} {2019})}\BibitemShut {NoStop}%
  \bibitem [{\citenamefont {Hornischer}\ \emph {et~al.}(2019)\citenamefont
    {Hornischer}, \citenamefont {Herminghaus},\ and\ \citenamefont
    {Mazza}}]{hornischer2019intelligence}%
    \BibitemOpen
    \bibfield  {author} {\bibinfo {author} {\bibfnamefont {H.}~\bibnamefont
    {Hornischer}}, \bibinfo {author} {\bibfnamefont {S.}~\bibnamefont
    {Herminghaus}},\ and\ \bibinfo {author} {\bibfnamefont {M.~G.}\ \bibnamefont
    {Mazza}},\ }\bibfield  {title} {\bibinfo {title} {Structural transition in
    the collective behavior of cognitive agents},\ }\href@noop {} {\bibfield
    {journal} {\bibinfo  {journal} {Scientific reports}\ }\textbf {\bibinfo
    {volume} {9}} (\bibinfo {year} {2019})}\BibitemShut {NoStop}%
  \bibitem [{Note1()}]{Note1}%
    \BibitemOpen
    \bibinfo {note} {It is not strictly self-consistent in the sense that the
    model for others (ballistic motion) is different to that of self (path
    entropy-maximisation). It is unclear whether such models can ever be made
    rigorously self-consistent or whether they then represent uncomputable
    functions \cite {turing_1937,Godel}}\BibitemShut {NoStop}%
  \bibitem [{\citenamefont {Ballerini}\ \emph
    {et~al.}(2008{\natexlab{b}})\citenamefont {Ballerini}, \citenamefont
    {Cabibbo}, \citenamefont {Candelier}, \citenamefont {Cavagna}, \citenamefont
    {Cisbani}, \citenamefont {Giardina}, \citenamefont {Orlandi}, \citenamefont
    {Parisi}, \citenamefont {Procaccini}, \citenamefont {Viale},\ and\
    \citenamefont {Zdravkovic}}]{BALLERINI2008201}%
    \BibitemOpen
    \bibfield  {author} {\bibinfo {author} {\bibfnamefont {M.}~\bibnamefont
    {Ballerini}}, \bibinfo {author} {\bibfnamefont {N.}~\bibnamefont {Cabibbo}},
    \bibinfo {author} {\bibfnamefont {R.}~\bibnamefont {Candelier}}, \bibinfo
    {author} {\bibfnamefont {A.}~\bibnamefont {Cavagna}}, \bibinfo {author}
    {\bibfnamefont {E.}~\bibnamefont {Cisbani}}, \bibinfo {author} {\bibfnamefont
    {I.}~\bibnamefont {Giardina}}, \bibinfo {author} {\bibfnamefont
    {A.}~\bibnamefont {Orlandi}}, \bibinfo {author} {\bibfnamefont
    {G.}~\bibnamefont {Parisi}}, \bibinfo {author} {\bibfnamefont
    {A.}~\bibnamefont {Procaccini}}, \bibinfo {author} {\bibfnamefont
    {M.}~\bibnamefont {Viale}},\ and\ \bibinfo {author} {\bibfnamefont
    {V.}~\bibnamefont {Zdravkovic}},\ }\bibfield  {title} {\bibinfo {title}
    {Empirical investigation of starling flocks: a benchmark study in collective
    animal behaviour},\ }\href
    {https://doi.org/https://doi.org/10.1016/j.anbehav.2008.02.004} {\bibfield
    {journal} {\bibinfo  {journal} {Animal Behaviour}\ }\textbf {\bibinfo
    {volume} {76}},\ \bibinfo {pages} {201} (\bibinfo {year}
    {2008}{\natexlab{b}})}\BibitemShut {NoStop}%
  \bibitem [{\citenamefont {Hemelrijk}\ and\ \citenamefont
    {Hildenbrandt}(2011)}]{HemelrijkPLOSONE2011}%
    \BibitemOpen
    \bibfield  {author} {\bibinfo {author} {\bibfnamefont {C.~K.}\ \bibnamefont
    {Hemelrijk}}\ and\ \bibinfo {author} {\bibfnamefont {H.}~\bibnamefont
    {Hildenbrandt}},\ }\bibfield  {title} {\bibinfo {title} {Some causes of the
    variable shape of flocks of birds},\ }\href
    {https://doi.org/10.1371/journal.pone.0022479} {\bibfield  {journal}
    {\bibinfo  {journal} {PLOS ONE}\ }\textbf {\bibinfo {volume} {6}},\ \bibinfo
    {pages} {1} (\bibinfo {year} {2011})}\BibitemShut {NoStop}%
  \bibitem [{\citenamefont {Ester}\ \emph {et~al.}(1996)\citenamefont {Ester},
    \citenamefont {Kriegel}, \citenamefont {Sander},\ and\ \citenamefont
    {Xu}}]{ester1996density}%
    \BibitemOpen
    \bibfield  {author} {\bibinfo {author} {\bibfnamefont {M.}~\bibnamefont
    {Ester}}, \bibinfo {author} {\bibfnamefont {H.-P.}\ \bibnamefont {Kriegel}},
    \bibinfo {author} {\bibfnamefont {J.}~\bibnamefont {Sander}},\ and\ \bibinfo
    {author} {\bibfnamefont {X.}~\bibnamefont {Xu}},\ }\bibfield  {title}
    {\bibinfo {title} {A density-based algorithm for discovering clusters in
    large spatial databases with noise},\ }in\ \href@noop {} {\emph {\bibinfo
    {booktitle} {Proceedings of the Second International Conference on Knowledge
    Discovery and Data Mining}}},\ \bibinfo {series and number} {KDD'96}\
    (\bibinfo  {publisher} {AAAI Press},\ \bibinfo {year} {1996})\ p.\ \bibinfo
    {pages} {226–231}\BibitemShut {NoStop}%
  \bibitem [{\citenamefont {Turing}(1937)}]{turing_1937}%
    \BibitemOpen
    \bibfield  {author} {\bibinfo {author} {\bibfnamefont {A.~M.}\ \bibnamefont
    {Turing}},\ }\bibfield  {title} {\bibinfo {title} {Computability and
    $\lambda$-definability},\ }\href {https://doi.org/10.2307/2268280} {\bibfield
     {journal} {\bibinfo  {journal} {Journal of Symbolic Logic}\ }\textbf
    {\bibinfo {volume} {2}},\ \bibinfo {pages} {153} (\bibinfo {year}
    {1937})}\BibitemShut {NoStop}%
  \bibitem [{\citenamefont {G{\"o}del}(1931)}]{Godel}%
    \BibitemOpen
    \bibfield  {author} {\bibinfo {author} {\bibfnamefont {K.}~\bibnamefont
    {G{\"o}del}},\ }\bibfield  {title} {\bibinfo {title} {{\"U}ber formal
    unentscheidbare s{\"a}tze der principia mathematica und verwandter systeme
    i},\ }\href {https://doi.org/10.1007/BF01700692} {\bibfield  {journal}
    {\bibinfo  {journal} {Monatshefte f{\"u}r Mathematik und Physik}\ }\textbf
    {\bibinfo {volume} {38}},\ \bibinfo {pages} {173} (\bibinfo {year}
    {1931})}\BibitemShut {NoStop}%
  \end{thebibliography}
\end{document}